\documentclass[apj]{emulateapj}
\shorttitle{S0 EVOLUTION VS ENVIRONMENT}
\shortauthors{JUST ET AL.}
\newcommand{\tx}[1]{\textrm{#1}}

\newcommand{\kms}{km~$\tx{s}^{-1}$}

\newcommand{\fso}{{$f_{\rm S0}$}}
\newcommand{\fe}{$f_{\rm E}$}

\def\gtsim{\lower 2pt \hbox{$\, \buildrel {\scriptstyle >}\over  {\scriptstyle \sim}\,$}}
\def\ltsim{\lower 2pt \hbox{$\, \buildrel {\scriptstyle <}\over  {\scriptstyle \sim}\,$}}
\newcommand{\hisig}{$\sigma\gtsim 750$~km~s$^{-1}$}
\newcommand{\losig}{$\sigma\ltsim 750$~km~s$^{-1}$}

\journalinfo{The Astrophysical Journal, ???:??--??, 2010}
\slugcomment{Accepted 2010 Jan 12}

\begin{document}

\title{The Environmental Dependence of the Evolving S0 Fraction{*}}
\author{Dennis~W. Just\altaffilmark{1},
Dennis~Zaritsky\altaffilmark{1}, David~J.~Sand\altaffilmark{1,2,3},
Vandana Desai\altaffilmark{4,5}, Gregory~Rudnick\altaffilmark{6}}
\altaffiltext{*}{Observations reported here were obtained at the MMT
Observatory, a joint facility of the University of Arizona and the
Smithsonian Institution; and based on data collected at the Magellan
Telescope, which is operated by the Carnegie Observatories.}
\altaffiltext{1}{Steward Observatory, University of Arizona, 933 North
Cherry Avenue, Tucson, AZ 85721, USA}
\altaffiltext{2}{Harvard Center for Astrophysics and Las Cumbres
Observatory Global Telescope Network Fellow}
\altaffiltext{3}{Harvard-Smithsonian Center for Astrophysics, 60
Garden Street, Cambridge, MA 02138, USA}
\altaffiltext{4}{Division of Physics, Mathematics and Astronomy,
California Institute of Technology, Pasadena, CA 91125, USA}
\altaffiltext{5}{Spitzer Science Center, California Institute of
Technology, Pasadena, CA 91125, USA}
\altaffiltext{6}{The University of Kansas, Department of Physics and
Astronomy, Malott room 1082, 1251 Wescoe Hall Drive, Lawrence, KS
66045, USA}


\begin{abstract}
We reinvestigate the dramatic rise in the S0 fraction, \fso, within
clusters since $z \sim 0.5$. In particular, we focus on the role of
the global galaxy environment on \fso\ by compiling, either from our
own observations or the literature, robust line-of-sight velocity
dispersions, $\sigma's$, for a sample of galaxy groups and clusters at
$0.1 < z < 0.8$ that have uniformly determined, published
morphological fractions.  We find that the trend of \fso\ with
redshift is twice as strong for $\sigma<750$~km~s$^{-1}$ groups/poor
clusters than for higher-$\sigma$, rich clusters. From this result, we
infer that over this redshift range galaxy-galaxy interactions, which
are more effective in lower-$\sigma$ environments, are more
responsible for transforming spiral galaxies into S0's than
galaxy-environment processes, which are more effective in
higher-$\sigma$ environments. The rapid, recent growth of the S0
population in groups and poor clusters implies that large numbers of
progenitors exist in low-$\sigma$ systems at modest redshifts ($\sim
0.5$), where morphologies and internal kinematics are within the
measurement range of current technology.
\end{abstract}

\section{Introduction}

The fraction of galaxies morphologically classified as S0 (\fso)
increases by a factor of $\sim 3$ in galaxy groups and clusters over
the past $\sim 5$~Gyr, at the expense of the spiral fraction
\citep{Dressler97}. This evolution has generally been interpreted as
the result of the transformation of spirals into S0's within dense
environments (\cite{Dressler97,Fasano00}, hereafter F00;
\cite{Smith05,Postman05,Poggianti06,Desai07}, hereafter D07), although
the physical mechanism remains undetermined.  As highlighted by
\cite{Dressler80}, the relationship between morphologies and
environment can help distinguish between hypothesized formation
mechanisms for S0's.  As practiced, this effort involves tracing
galaxy populations as a function of environment \citep{Dressler80,
Postman84, Zabludoff98, Helsdon03}, increasingly at higher redshifts
\citep{Dressler97,Kautsch08, Wilman09}.  Those studies in turn have
produced the evidence for significant evolution of the S0 fraction
\citep{Dressler97}, but have not examined whether the rate of
evolution itself depends on environment.

We focus on the relationship between S0 evolution and the velocity
dispersion ($\sigma$) of the group or cluster that hosts the
S0's. Processes that are expected to operate best in lower-$\sigma$
environments, where the lower relative velocities between galaxies
allow them to interact more effectively, include mergers and
galaxy-galaxy interactions
\citep{Toomre72,Icke85,Lavery88,Byrd90,Mihos04}. Those expected to
work best in higher-$\sigma$ environments, either directly because of
the high velocities, the deeper potential implied by the high
velocities, or the higher density intracluster medium, include ram
pressure stripping \citep{Gunn72,Abadi99,Quilis00}, strangulation
\citep{Larson80,Bekki02}, and harassment \citep{Richstone76,Moore98}.

To investigate the dependence of \fso\ on environment, we return to
published morphological samples. We use published visual morphological
classifications as the indicator of galaxy type. Quantities related to
\fso, such as $B/T$ and color distributions, have also been used to
investigate such questions, but morphologies provide additional,
complementary information. In fact, various recent studies are
suggesting that morphological evolution is somewhat decoupled from the
evolution of the stellar population \citep{Poggianti06,Tran09}.
Morphologies are available across a significant range of redshifts and
velocity dispersions, and significant effort has been expended in
putting these on a common footing across redshift (F00; D07). We
compile an internally-consistent set of velocity dispersions,
recalculating the velocity dispersion using either previously
published individual galaxy redshifts or redshifts from our own
observations, to provide a measure of environment. Again, alternative
measurements of environment exist, for example X-ray luminosities
could have been used. However, X-ray measurements, particularly for
low-mass, high-redshift environments, are scarce and velocity
dispersions provide the most uniform and extensive data. Studies using
different measures of either galaxy type or environment are mixed. For
example X-ray luminosities correlate with $B/T$ at $z\sim0$
\citep{Balogh02} and with early-type fraction at $z>1$
\citep{Postman05}, but velocity dispersions correlate only weakly with
the fraction of red galaxies within the virial radius
\citep{Balogh04}. Apparently conflicting results such as these
highlight the importance of using consistent measurements of both
galaxy type and environment across redshift when investigating
evolution.

In \S\ref{sec:data}, we describe the two samples we chose to use, the
spectroscopic measurements we acquired in an attempt to obtain
velocity dispersions to complete the sample, and the calculation of a
consistent set of velocity dispersion measurements. In
\S\ref{sec:results}, we present our results, discuss their
implications in \S\ref{sec:discussion}, and summarize in
\S\ref{sec:conclusion}.  When computing the aperture size used for
calculating the velocity dispersion, we assume
$H_{0}=70$~\kms~Mpc$^{-1}$, $\Omega_{m}=0.3$, and
$\Omega_{\Lambda}=0.7$ (hereafter, the ``Lambda cosmology''). However,
for the aperture size within which galaxies are included in the
calculation of morphological fractions, $H_{0}=50$~\kms~Mpc$^{-1}$,
$\Omega_{m}=1$, and $\Omega_{\Lambda}=0$ (hereafter, the ``classic
cosmology'') is assumed.

\section{Data}\label{sec:data}
\subsection{Sample}

Morphological fractions can depend sensitively on the aperture within
which cluster members are classified and on the absolute magnitude to
which the classification is done. As such, it can be quite difficult,
and potentially misleading, to use classifications from disparate
sources.  D07 presented their own classification of a set of galaxies
and combined these with a set from the literature for which they were
able to closely match the classification procedure, the aperture used,
and the magnitude limit. Specifically, the sample presented in D07
consists of 23 galaxy clusters at $z\sim0.1$--0.5 drawn from the F00
sample and 10 clusters at $z\sim0.5$--0.8 drawn from EDisCS. The F00
sample in turn consists of nine clusters at $0.1\ltsim z \ltsim 0.3$
added by the authors themselves, five clusters at $0.15\ltsim z \ltsim
0.3$ that either appeared in \cite{Couch98} or were classified in a
manner consistent with that study, and nine clusters at $0.3\ltsim z
\ltsim 0.5$ from the MORPHS study \citep{Dressler97,Smail97}, all of
which were classified in a consistent manner.  D07 used the F00
procedure when classifying galaxies to minimize systematic differences
between the two samples; in particular, the five authors who did the
morphological classification also reclassified the highest redshift
clusters of F00 (from $0.3<z<0.5$), following the same procedure as
the original authors \citep{Smail97}, and found good agreement.

Errors on the morphological fractions for those from the ESO Distant
Cluster Survey \citep[EDisCS;][]{White05} were computed using the
method of \cite{Gehrels86}. The situation is somewhat more complicated
for the F00 morphological fractions.  We calculate the uncertainties
using the Gehrels method, but some of the necessary information, such
as the various correction and completeness factors, are not available
and we infer them indirectly from the data provided by F00. To test
the sensitivity of our results to the uncertainties, we also do all
the analysis described subsequently using the quoted uncertainties in
F00, which were not calculated using the Gehrels method.  None of the
results (including the statistical significances quoted) change
sufficiently between the two approaches to alter any of our
conclusions.  To directly compare their results to F00, who present
morphological fractions for non-uniform apertures that correspond to
apertures of radii spanning from $\sim500$ to 700 kpc, D07 used the
classic cosmology to measure morphological fractions within fixed
600~kpc radius apertures for the EDisCS clusters.  This selection of a
fixed physical aperture attempts to best match, on average, the F00
measurements, which are for a range of apertures. However, D07
demonstrated that a choice of aperture that scales with $R_{200}$
($0.6R_{200}$) results in \fso\ values that are in all cases within
the uncertainty estimates. Lastly, regarding the magnitude limit, D07
classify galaxies down to a fixed absolute magnitude across the
redshift range, chosen to match the F00 classification procedure,
assuming the rest-frame colors and $I$-band magnitude of an elliptical
galaxy (details provided in D07). Applying the incorrect cosmology
(i.e. classic rather than Lambda cosmology) results in differential
magnitude limits across the redshift range from 0.2 to 0.8 of a few
tenths of a magnitude, comparable to the uncertainties in the observed
magnitudes themselves and therefore not expected to have a noticeable
effect.

A sample of $z\sim 1$ clusters with morphological classifications and
redshift measurements from \cite{Postman05} also appear in
D07. However, those morphological fractions were not explicitly
matched to those of F00 (i.e., by taking steps to minimize systematic
differences in classification, such as those stated above) and,
therefore, we exclude these clusters to avoid any possible confusion
in the interpretation of our results. Including these clusters does
not alter our main results.

\begin{deluxetable*}{lcccccc}
\tablecolumns{7}
\tablenum{1} \tabletypesize{\scriptsize} \tablewidth{0pt}
\tablecaption{Log of Observations} \tablehead{ \colhead{} & \colhead{}
& \colhead{} & \colhead{} & \colhead{Total Redshifts} &
\colhead{Cluster Redshifts} & \colhead{} \\ \colhead{Cluster} &
\colhead{Date} & \colhead{Telescope} & \colhead{Instrument} &
\colhead{Measured} & \colhead{Measured} & \colhead{Notes} } \startdata
Abell~951 & 2007 Nov & MMT & Hectospec & 23 & 19 & \nodata \\
Abell~2658 & 2007 Oct & MMT & Hectospec & 146 & 41 & \nodata \\
Abell~1952 & 2008 Mar & MMT & Hectospec & 131 & 46 & \nodata \\
Abell~2192 & 2008 Mar & MMT & Hectospec & 100 & 13 & \nodata \\
Abell~1643 & 2008 Mar & MMT & Hectospec & \nodata & \nodata & Lost due
to weather. \\ Abell~1878 & 2008 Apr & MMT & Hectospec & \nodata &
\nodata & Lost due to weather. \\ & 2008 Jun & Magellan & IMACS & 25 &
18 & \nodata \\ Cl0054$-$27 & 2008 Jun & Magellan & IMACS & \nodata &
\nodata & Lost due to weather. \\ Abell~3330 & 2008 Sep & Magellan &
IMACS & \nodata & \nodata & Lost due to weather. \\ Cl0413$-$6559 &
2008 Sep & Magellan & IMACS & \nodata & \nodata & Lost due to
weather. \\ \enddata
\end{deluxetable*}

\subsection{New and Archival Redshifts}
Of the 33 galaxy clusters and groups from the combined sample of F00
and EDisCS, seven ($\sim 20\%$; all from F00) do not have previously
published velocity dispersion measurements. All of these clusters are
at $z<0.25$, where less than half of the clusters have velocity
dispersion measurements. This important part of parameter space drives
much of the \fso-$z$ trend observed in F00. Although several of these
clusters have enough individual galaxy redshifts available in the
literature with which to calculate a reliable velocity dispersion
\citep[$\gtsim 10$, see][]{Beers90}, we still targeted them for
observation because a higher number of redshifts allows us to
calculate a more robust velocity dispersion. We targeted six clusters
(Abell~951, Abell~1643, Abell~1878, Abell~1952, Abell~2192, and
Abell~2658) using Hectospec \citep{Fabricant05} on the MMT between
2007 November to 2008 April. We observed each cluster for a total of
$30$--$60$~minutes and measured redshifts using the {\sc iraf} task
{\sl rvsao}. We used HSRED \citep[e.g., \S3.2 of][]{Papovich06} for
the Hectospec data reduction. We also targeted four clusters
(Abell~1878, Abell~3330, Cl0054$-$27, and Cl0413$-$6559) using the
Inamori-Magellan Areal Camera and Spectrograph
\citep[IMACS;][]{Bigelow98} on the Magellan Baade telescope during two
observation runs in 2008 June and 2008 September. IMACS data was
reduced using the COSMOS package\footnote{The Carnegie Observatories
System for Multiobject Spectroscopy was created by A. Oemler,
K. Clardy, D. Kelson, and G. Walth. See
http://www.ociw.edu/Code/cosmos.}, following standard reduction
procedures. Based on our comparison of 15 objects for which previous
redshift measurements exist, we calculate that our velocity
measurement uncertainty is 86 km sec$^{-1}$. This is a conservative
estimate in that we assign the entire difference between our
measurements and the published ones to ourselves.

A log of the observations of the clusters is presented in Table~1. The
target galaxies are selected from the NASA Extragalactic Database
(NED) and so there is no uniform selection criteria. We prioritize
what appear to be early-type galaxies and use whatever other
information is in NED to maximize our return on cluster members, but
given the heterogeneity of the source material the target sample is
ill-defined.  Furthermore, as with all multiobject spectroscopy, the
effective selection is complicated by fiber/slit allocation algorithms
and then by the intrinsic spectrum of an object. In detail, such
biases can lead to differences in measured velocity dispersions due to
differences in the dispersions of different morphogical types within a
cluster \citep[cf.][]{zab}, but here we use the velocity dispersions
only as a rough ranking of environment and are not interested in
differences at the $\sim 10$\% level. Both of these spectrographs
provide large ($> 24$ arcmin) wide fields-of-view, so the galaxies
sample the dynamics well beyond the cluster core. These observations
provide enough redshifts for all but one cluster (Abell~1643 from the
F00 sample, which was observed during poor weather) to measure the
velocity dispersions for nearly the full sample (32/33 clusters). The
other clusters lost due to weather had enough redshifts to reliably
measure the velocity dispersion. In the analyses that follow, only
these 32 clusters are included.

\begin{deluxetable*}{lccccccrcccrc}
\tablecolumns{13}
\tablenum{2} \tabletypesize{\scriptsize} \tablewidth{0pt}
\tablecaption{Main Properties of the Sample} \tablehead{
\colhead{Name} & \colhead{$z$} & \colhead{$\sigma$} & \colhead{$R_{\rm
200}$} & \colhead{$N_{\rm iter}$} & \colhead{$N_{\rm mem}$} &
\colhead{$f_{\rm E}$} & \colhead{$f_{\rm S0}$} & \colhead{$f_{\rm S}$}
& \colhead{$f_{\rm E+S0}$} & \colhead{Sample} \\ \colhead{(1)} &
\colhead{(2)} & \colhead{(3)} & \colhead{(4)} & \colhead{(5)} &
\colhead{(6)} & \colhead{(7)} & \colhead{(8)} & \colhead{(9)} &
\colhead{(10)} & \colhead{(11)} } \startdata A3330 & 0.091 &
732$^{+237}_{-82}$ & 1.73 &\nodata& 9 & 0.307$^{+0.089}_{-0.070}$ &
0.501$^{+0.083}_{-0.084}$ & 0.193$^{+0.085}_{-0.055}$ &
0.807$^{+0.056}_{-0.085}$ & 1 \\ A389 & 0.116 & 662$^{+175}_{-130}$ &
1.55 & 3 & 40 & 0.353$^{+0.094}_{-0.088}$ & 0.629$^{+0.099}_{-0.087}$
& 0.019$^{+0.070}_{-0.014}$ & 0.981$^{+0.014}_{-0.070}$ & 1 \\ A951* &
0.143 & 537$^{+128}_{-66}$ & 1.24 & 4 & 23 & 0.313$^{+0.127}_{-0.095}$
& 0.649$^{+0.098}_{-0.129}$ & 0.038$^{+0.096}_{-0.031}$ &
0.962$^{+0.031}_{-0.096}$ & 1 \\ A2218 & 0.171 & 1520$^{+112}_{-74}$ &
3.45 & 1 & 98 & 0.437$^{+0.092}_{-0.085}$ & 0.240$^{+0.090}_{-0.067}$
& 0.324$^{+0.083}_{-0.085}$ & 0.677$^{+0.085}_{-0.083}$ & 1 \\ A1689 &
0.181 & 1876$^{+98}_{-71}$ & 4.24 & 1 & 206 &
0.363$^{+0.063}_{-0.051}$ & 0.363$^{+0.063}_{-0.051}$ &
0.274$^{+0.059}_{-0.048}$ & 0.726$^{+0.048}_{-0.059}$ & 1 \\ A2658* &
0.185 & 498$^{+99}_{-58}$ & 1.12 &\nodata& 15 &
0.491$^{+0.121}_{-0.152}$ & 0.410$^{+0.152}_{-0.119}$ &
0.099$^{+0.130}_{-0.062}$ & 0.901$^{+0.062}_{-0.130}$ & 1 \\ A2192* &
0.187 & 635$^{+139}_{-112}$ & 1.43 &\nodata& 16 &
0.287$^{+0.085}_{-0.076}$ & 0.511$^{+0.077}_{-0.099}$ &
0.202$^{+0.095}_{-0.054}$ & 0.798$^{+0.054}_{-0.095}$ & 1 \\ A1643 &
0.198 & \nodata &\nodata&\nodata&\nodata& 0.242$^{+0.070}_{-0.073}$ &
0.476$^{+0.075}_{-0.090}$ & 0.282$^{+0.075}_{-0.075}$ &
0.718$^{+0.075}_{-0.075}$ & 1 \\ A1878* & 0.222\tablenotemark{a} &
828$^{+280}_{-135}$ & 1.83 & 1 & 13 & 0.364$^{+0.106}_{-0.083}$ &
0.282$^{+0.070}_{-0.104}$ & 0.354$^{+0.116}_{-0.073}$ &
0.646$^{+0.073}_{-0.116}$ & 1 \\ A2111*\tablenotemark{b} & 0.229 &
1129$^{+121}_{-80}$ & 2.49 & 2 & 80 & 0.465$^{+0.066}_{-0.067}$ &
0.336$^{+0.064}_{-0.063}$ & 0.200$^{+0.064}_{-0.047}$ &
0.800$^{+0.047}_{-0.064}$ & 1 \\ A1952* & 0.248 & 718$^{+293}_{-209}$
& 1.57 & 1 & 18 & 0.413$^{+0.078}_{-0.078}$ &
0.380$^{+0.072}_{-0.081}$ & 0.207$^{+0.082}_{-0.052}$ &
0.793$^{+0.052}_{-0.082}$ & 1 \\ AC118 & 0.308 & 1748$^{+99}_{-139}$ &
3.69 & 1 & 83 & 0.246$^{+0.061}_{-0.053}$ & 0.527$^{+0.064}_{-0.064}$
& 0.227$^{+0.062}_{-0.049}$ & 0.773$^{+0.049}_{-0.062}$ & 1 \\ AC103 &
0.311 & 965$^{+132}_{-81}$ & 2.03 & 1 & 55 & 0.301$^{+0.078}_{-0.071}$
& 0.313$^{+0.086}_{-0.064}$ & 0.386$^{+0.075}_{-0.081}$ &
0.614$^{+0.081}_{-0.075}$ & 1 \\ AC114 & 0.315 & 1889$^{+81}_{-74}$ &
3.98 & 1 & 196 & 0.223$^{+0.049}_{-0.051}$ & 0.318$^{+0.061}_{-0.050}$
& 0.459$^{+0.060}_{-0.058}$ & 0.541$^{+0.058}_{-0.060}$ & 1 \\
Cl1446$+$2619 & 0.370 & 1397$^{+287}_{-218}$ & 2.85 & 2 & 20 &
0.338$^{+0.082}_{-0.070}$ & 0.248$^{+0.074}_{-0.068}$ &
0.415$^{+0.086}_{-0.072}$ & 0.585$^{+0.072}_{-0.086}$ & 1 \\
Cl0024$+$1652 & 0.391 & 764$^{+40}_{-50}$ & 1.54 & 2 & 235 &
0.348$^{+0.084}_{-0.076}$ & 0.227$^{+0.075}_{-0.070}$ &
0.426$^{+0.082}_{-0.085}$ & 0.574$^{+0.085}_{-0.082}$ & 1 \\
Cl0939$+$4713 & 0.405 & 1331$^{+96}_{-109}$ & 2.65 & 1 & 72 &
0.250$^{+0.095}_{-0.068}$ & 0.257$^{+0.097}_{-0.070}$ &
0.493$^{+0.100}_{-0.086}$ & 0.507$^{+0.086}_{-0.100}$ & 1 \\
Cl0303$+$1706 & 0.418 & 769$^{+120}_{-94}$ & 1.52 & 2 & 56 &
0.227$^{+0.084}_{-0.072}$ & 0.126$^{+0.075}_{-0.054}$ &
0.647$^{+0.085}_{-0.088}$ & 0.353$^{+0.088}_{-0.085}$ & 1 \\ 3C295 &
0.461 & 1907$^{+142}_{-205}$ & 3.69 & 1 & 32 &
0.463$^{+0.093}_{-0.101}$ & 0.197$^{+0.095}_{-0.067}$ &
0.341$^{+0.100}_{-0.086}$ & 0.659$^{+0.086}_{-0.100}$ & 1 \\
Cl0412$-$6559 & 0.510 & 626$^{+210}_{-179}$ & 1.17 & 1 & 19 &
0.347$^{+0.089}_{-0.089}$ & 0.090$^{+0.064}_{-0.053}$ &
0.564$^{+0.080}_{-0.105}$ & 0.437$^{+0.105}_{-0.080}$ & 1 \\
Cl1601$+$42 & 0.539 & 749$^{+97}_{-76}$ & 1.38 & 1 & 55 &
0.509$^{+0.064}_{-0.068}$ & 0.165$^{+0.061}_{-0.042}$ &
0.326$^{+0.068}_{-0.058}$ & 0.674$^{+0.058}_{-0.068}$ & 1 \\
Cl0016$+$16 & 0.545 & 1307$^{+112}_{-113}$ & 2.41 & 2 & 99 &
0.502$^{+0.076}_{-0.080}$ & 0.208$^{+0.076}_{-0.055}$ &
0.291$^{+0.074}_{-0.069}$ & 0.709$^{+0.069}_{-0.074}$ & 1 \\
Cl0054$-$27 & 0.560 & 700$^{+284}_{-254}$ & 1.28 & 2 & 17 &
0.310$^{+0.087}_{-0.077}$ & 0.246$^{+0.084}_{-0.073}$ &
0.444$^{+0.085}_{-0.092}$ & 0.556$^{+0.092}_{-0.085}$ & 1 \\
Cl1138$-$1133 & 0.480 & 746$^{+96}_{-79}$ & 1.43 & 1 & 49 &
0.305$^{+0.164}_{-0.120}$ & 0.095$^{+0.113}_{-0.084}$ &
0.600$^{+0.145}_{-0.162}$ & 0.400$^{+0.162}_{-0.145}$ & 2 \\
Cl1232$-$1250 & 0.541 & 1171$^{+155}_{-70}$ & 2.16 & 1 & 54 &
0.350$^{+0.040}_{-0.040}$ & 0.170$^{+0.030}_{-0.030}$ &
0.470$^{+0.040}_{-0.040}$ & 0.530$^{+0.040}_{-0.040}$ & 2 \\
Cl1037$-$1243 & 0.578 & 344$^{+73}_{-64}$ & 0.58 & 1 & 16 &
0.281$^{+0.124}_{-0.146}$ & 0.000$^{+0.109}_{-0.000}$ &
0.625$^{+0.138}_{-0.156}$ & 0.281$^{+0.124}_{-0.146}$ & 2 \\
Cl1227$-$1138 & 0.636 & 584$^{+93}_{-70}$ & 0.64 &\nodata& 22 &
0.290$^{+0.165}_{-0.136}$ & 0.146$^{+0.157}_{-0.095}$ &
0.394$^{+0.167}_{-0.164}$ & 0.436$^{+0.160}_{-0.162}$ & 2 \\
Cl1054$-$1146 & 0.697 & 603$^{+170}_{-140}$ & 1.01 & 2 & 33 &
0.245$^{+0.071}_{-0.069}$ & 0.000$^{+0.036}_{-0.000}$ &
0.755$^{+0.069}_{-0.071}$ & 0.245$^{+0.071}_{-0.069}$ & 2 \\
Cl1103$-$1245b & 0.703 & 235$^{+203}_{-86}$ & 0.39 &\nodata& 9 &
0.250$^{+0.120}_{-0.080}$ & 0.000$^{+0.070}_{-0.000}$ &
0.750$^{+0.080}_{-0.120}$ & 0.250$^{+0.120}_{-0.080}$ & 2 \\
Cl1040$-$1155 & 0.704 & 535$^{+89}_{-71}$ & 0.89 & 2 & 15 &
0.377$^{+0.136}_{-0.116}$ & 0.066$^{+0.093}_{-0.058}$ &
0.419$^{+0.141}_{-0.115}$ & 0.444$^{+0.141}_{-0.123}$ & 2 \\
Cl1054$-$1245 & 0.750 & 570$^{+141}_{-103}$ & 0.93 & 2 & 22 &
0.300$^{+0.107}_{-0.090}$ & 0.267$^{+0.104}_{-0.087}$ &
0.433$^{+0.108}_{-0.102}$ & 0.567$^{+0.102}_{-0.108}$ & 2 \\
Cl1354$-$1230 & 0.762 & 732$^{+233}_{-48}$ & 1.18 & 1 & 21 &
0.170$^{+0.070}_{-0.050}$ & 0.290$^{+0.070}_{-0.060}$ &
0.550$^{+0.080}_{-0.070}$ & 0.450$^{+0.070}_{-0.080}$ & 2 \\
Cl1216$-$1201 & 0.794 & 1066$^{+82}_{-84}$ & 1.69 & 1 & 67 &
0.490$^{+0.030}_{-0.020}$ & 0.220$^{+0.020}_{-0.020}$ &
0.270$^{+0.020}_{-0.020}$ & 0.710$^{+0.020}_{-0.020}$ & 2 \\ \enddata
\tablecomments{(1) Cluster Name. An asterisk (*) denotes a cluster
with new data; (2) Redshift; (3) Velocity Dispersion in units of
km~s$^{-1}$; (4) Virial Radius in units of Mpc; (5) Number of
iterations until convergence, see \S2.3; (6) Number of redshifts
ultimately used in calculating the value in Column~3; (7) Fraction of
Elliptical galaxies; (8) Fraction of S0 galaxies; (9) Fraction of
Spiral galaxies; (10) Fraction of Elliptical+S0 galaxies; (11) Sample,
1-Fasano et al. (2000), 2-EDiscS} \tablenotetext{a}{This redshift is
different than that which appears in F00, who use $z=0.254$. The
origin of the discrepancy can be traced back to Sandage, Kristian, \&
Westphal (1976), where two potential redshifts for the cluster are
given at $z=0.222$ and $z=0.254$. The lower value was assumed to be
foreground, so the latter value was adopted in later studies. However,
with our newly measured redshifts of 18 galaxies near the cluster
position that are within $\pm 0.015$ of the lower value and only 2
that are within $\pm 0.015$ of the higher value, we adopt $z=0.222$ as
the cluster redshift.}  \tablenotetext{b}{While no new redshifts have
been measured for this cluster, it's velocity dispersion has not been
published as far as the authors know, and is presented for the first
time here.}
\end{deluxetable*}

In all, we present new redshift measurements for five clusters (four
from Hectospec observations and one from IMACS observations). Although
this is a small number of clusters relative to the entire sample, they
lie in the region of parameter space responsible for much of the S0
evolution (i.e., low-$z$, high-\fso). In addition to these new
redshift measurements, we took advantage of the large number of
previously-measured redshifts available in the literature. These
redshifts came from various studies, and we used NED to search for and
select the data. This provides improved velocity dispersion
measurements for many of the clusters.

\subsection{Velocity Dispersion Measurements}
We calculate velocity dispersions for the entire sample, including
those with previously measured velocity dispersions, so that all
measurements for the velocity dispersion are calculated using the same
method.  We now describe our procedure for evaluating the velocity
dispersion, including our iterative procedure to define an
aperture. In the end, we find that the velocity dispersions have only
a slight dependence on the aperture as long as the aperture is a
significant fraction of the virial radius.

Starting with both the literature and newly-measured redshifts, we
include only those galaxies within 3~Mpc of the cluster center in our
initial estimate of the velocity dispersion, although we do not always
have spectroscopic redshifts out to that radius.  The cluster center
is as defined in the previous studies and remains unchanged through
our procedure. Because of the small number of spectroscopic members in
most of these clusters and the nature of the iterative procedure, we
use the initial center, which is often defined either by X-ray
contours, brightest cluster galaxy, or weak lensing contours rather
than from the galaxy population centroid.  Following
\cite{Halliday04}, we also apply a redshift cut of $\Delta z=0.015$
about the redshift of the cluster. Only redshifts from the literature
with quoted errors $\ltsim 0.01$ are included; a difference in
redshift of $0.01$ corresponds to $3000$~km~s$^{-1}$, which is much
larger than the velocity dispersion itself for even our richest
clusters. We use the biweight statistic of \cite{Beers90} to calculate
the value of $\sigma$, which gives robust velocity dispersion
measurements with as few as $\sim 10$ galaxy redshift measurements.
The velocity dispersions are corrected to be rest-frame velocity
dispersions.  Regarding our choice of initial aperture, we find that
varying it within the range $\sim 1.5$--3~Mpc affects the velocity
dispersion by $\ltsim 10\%$ for all our clusters, most often $\ltsim
5\%$.  In fact, the velocity dispersion calculated within any aperture
varying from $\sim 1.5$--3~Mpc (when not implementing our iterative
aperture scheme outlined below) changes by $\ltsim 15\%$ for all our
clusters except Abell~951 and Abell~2658, whose velocity dispersions
change by $\sim 50\%$ within that range. After calculating the
velocity dispersion, $3\sigma$ outliers are rejected and the process
iterated until no outliers remain \citep[see \S5.2
of][]{Halliday04}. This value of $\sigma$ is then used to calculate an
estimated virial radius, $R_{200}$, using Equation~(5) of
\cite*{Finn04}:
\begin{equation}
R_{200}=1.73\frac{\sigma}{\rm 1000~km~s^{-1}}[\Omega_\Lambda +
\Omega_0(1+z)^3]^{-1/2} ~h^{-1}_{100}~ {\rm Mpc}.
\end{equation}
A new cut is applied at $R_{200}$, and the process iterated until
convergence. Sometimes $R_{200}$ is greater than 3~Mpc, resulting in
more redshifts being included in the later iterations. The main
properties of our 32 cluster sample, including these new velocity
dispersion measurements, appears in Table~2. The values for $R_{200}$,
the number of iterations until convergence, $N_{\rm iter}$, and the
number of redshifts used in the final iteration, $N_{\rm mem}$, appear
in Columns 4, 5, and 6, of Table~2, respectively. For five of the
clusters, Abell~3330, Abell~2658, Abell~2192, Cl1103$-$1245b, and
Cl1227$-$1138, this process of iteration removes galaxy redshifts
until there are too few ($\ltsim 10$) to reliably calculate a velocity
dispersion.  For these systems, the velocity dispersion is calculated
using a fixed 3~Mpc cut, and the $R_{200}$ that appears in Table~2 is
calculated from Equation~(1) using the velocity dispersion obtained
with that aperture. We estimate the 1$\sigma$ errors by selecting
random subsamples of the data from which to evaluate the velocity
dispersion.

For three of the clusters, Abell~1952, Cl0024$+$1652 (both part of the
F00 subsample), and Cl1037$-$1243 (part of the EDisCS subsample),
there is clear\footnote{For Cl1037$-$1243, the substructure only
becomes obvious after the first iteration. Two galaxies located
$2^{\prime\prime}$ apart on the sky have velocities of $\approx -1500$
and $-2000$~km~s$^{-1}$ relative to the cluster. Due to the relatively
few galaxies in the cluster (16), these two galaxies change the
velocity dispersion from $\approx 300$ to $650$~km~s$^{-1}$ when they
are included (such that they are then not excluded in the 3$\sigma$
clipping). Inspection of the histogram leads us to believe the former
value is more accurate, although adopting the latter value does not
significantly change our results.} evidence of substructure in their
phase-space plots. We remove by hand the galaxies belonging to these
subgroups when calculating the velocity dispersion for the three
clusters. Aside from this step, the velocity dispersion is calculated
using the same procedure outlined above.

We present velocity histograms for the clusters in Figure~\ref{hist}
(placed at the end of the paper).  The bin size is set to one-third
the velocity dispersion, and the redshifts plotted are those that
remain after the various cuts/iterations in the calculation (see
above). Overplotted on each panel is a Gaussian with the measured
velocity dispersion, normalized to the area of the histogram. Our
newly calculated velocity dispersions are in good agreement with those
previously measured for the EDisCS clusters
\citep{Halliday04,MilvangJensen08}, but tend to give larger values for
some of the $\sigma>1000$~km~s$^{-1}$ F00 clusters (see D07, and
references therein). This discrepancy is not due to aperture-size
effects, but more likely from the different methods employed in
calculating the velocity dispersion. Although the velocity dispersion
was calculated using the Lambda cosmology, while the morphological
fractions were calculated within an aperture defined by the classic
cosmology, we find that the value of $\sigma$ is fairly insensitive to
aperture size (see above).

Lastly, we address the impact of observational uncertainties on our
measured velocity dispersions. As mentioned previously, comparison of
our redshift measurements with those in the literature suggests a
single measurement uncertainty of 86 km s$^{-1}$. This is likely to be
a significant overestimate for most systems, but we use this value to
estimate the impact on the dispersions. If we simply add random
velocities using a Gaussian with this $\sigma$ to an intrinsic
Gaussian of width commensurate to the line-of-sight velocity
distribution of a specific group and cluster, we find that even in for
our lowest velocity dispersion system (Cl1102-1245b) the observational
errors inflate the dispersion by less than 20 km~s$^{-1}$.  This
uncertainty is in all cases significantly less than the quoted errors
on the velocity dispersion and does not affect our results.

\section{Results}\label{sec:results}

We explore the environmental dependence (characterized by velocity
dispersion) of the apparent evolution of \fso\ with redshift
(Figure~\ref{fs0vz}). Our sample spans a range of dispersions from
that typical of groups ($\sim 200-500$~km~s$^{-1}$) to poor clusters
($\sim 500-750$~km~s$^{-1}$) to rich clusters ($\gtsim
750$~km~s$^{-1}$). Although there is no strict rule for what velocity
dispersion constitutes a group versus a cluster, in what follows we
use the above convention.

\subsection{Analysis of the Full Sample}
We begin by determining whether a relationship between \fso\ and
environment (velocity dispersion) exists across the full redshift
range. Due to the selection criteria for the F00 sample (clusters were
selected based on being ``well-studied"), it is possible that some
unappreciated selection bias manifests itself as a correlation between
\fso\ and $z$. Figure~6 of D07 shows a weak trend between \fso\ and
$\sigma$, although they were limited to the subset of F00 clusters
with dispersion measurements and as we have noted earlier some of the
most interesting clusters were missing such measurements.

\begin{figure}
\epsscale{1.0}
\plotone{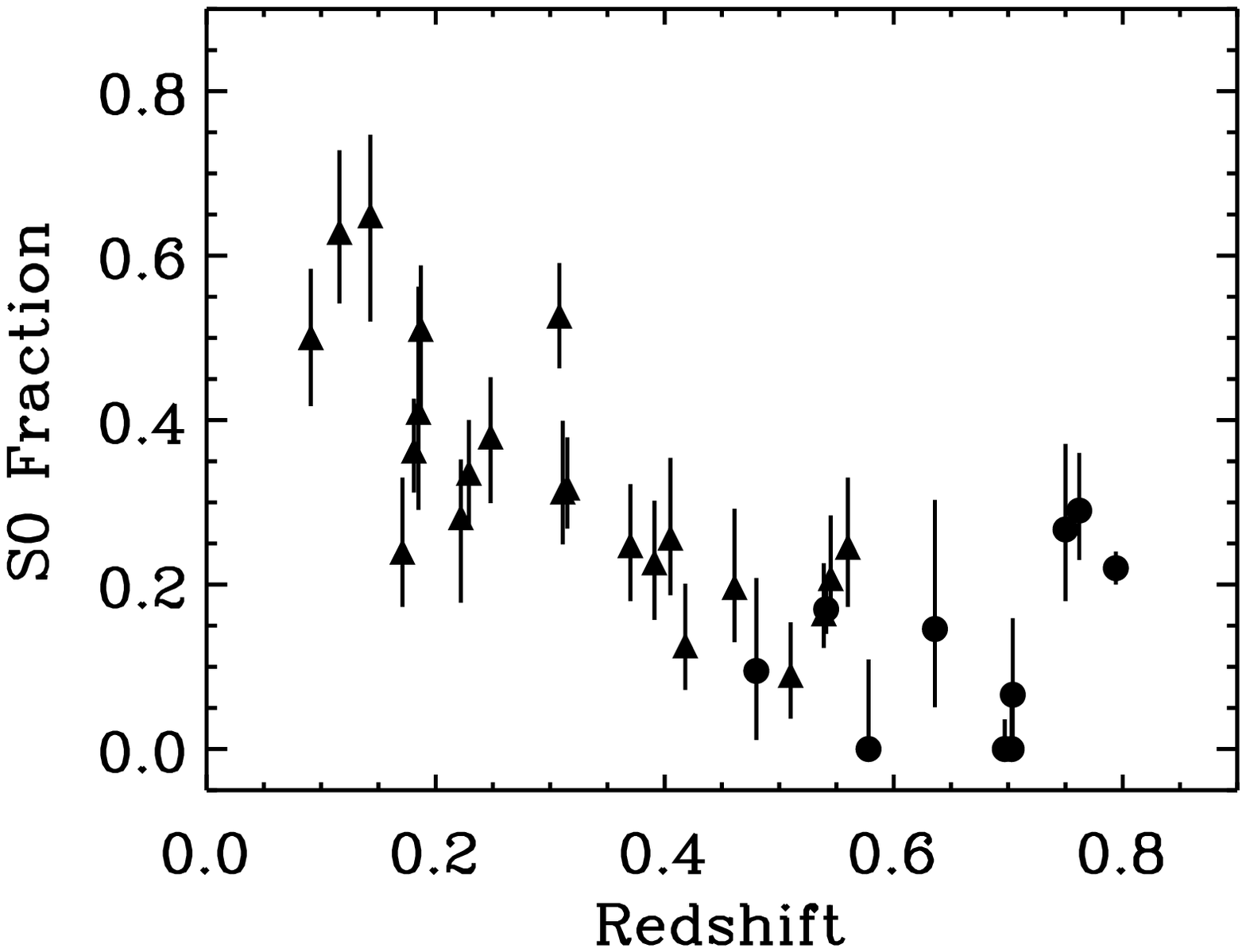}
\figurenum{2}
\caption{\label{fs0vz} S0 Fraction (\fso) plotted against
redshift. Triangles represent F00 systems, while circles represent
EDisCS systems.}
\end{figure}

In Figure~\ref{fs0vsig}, we present \fso\ plotted against velocity
dispersion for our sample. Although the \fso-$z$ trend in
Figure~\ref{fs0vz} appears much stronger than any trend between \fso\
and $\sigma$ in Figure~\ref{fs0vsig}, we quantify which is the more
dominant with a partial correlation analysis. The partial correlation
coefficient $\rho$,
\begin{equation}
\rho = {r_{A,B} - r_{A,C}r_{B,C} \over
\sqrt{(1-r_{A,C}^2)(1-r_{B,C}^2)}},
\end{equation}
is useful for disentangling the interdependence between three
variables (A, B, and C), where one wants to account for the influence
of the third variable (C) on the correlation of the first two.  It is
normalized to $+1$ for a perfect correlation, $0$ for no correlation,
and $-1$ for a perfect anticorrelation between A and B after
accounting for C. However, the distribution of $\rho$ does not
approximate a normal distribution, so we follow the work of
\cite{Kendall77} in using a statistic $Z_{B,C}$, where B is the
dependent variable and C is the controlled variable. $Z_{B,C}$ is
defined as

\begin{equation}
Z_{B,C} = \frac{1}{2}{\rm ln}\frac{(1+\rho)}{(1-\rho)}.
\end{equation}
with a variance $\sigma_Z^2=1/(N-2)$, where $N$ is the number of data
points. The more positive (negative) the value of $Z_{B,C}$ the
stronger the correlation (anticorrelation). We treat $z$ and $\sigma$
as the independent and controlled variable, and then vice-versa. We
find a stronger correlation for \fso\ with redshift than with
$\sigma$, with $Z_{z,\sigma}=-0.91\pm0.18$ and
$Z_{\sigma,z}=-0.02\pm0.18$.

\subsection{Analysis of Groups vs. Clusters}

\begin{figure}
\epsscale{1.0}
\plotone{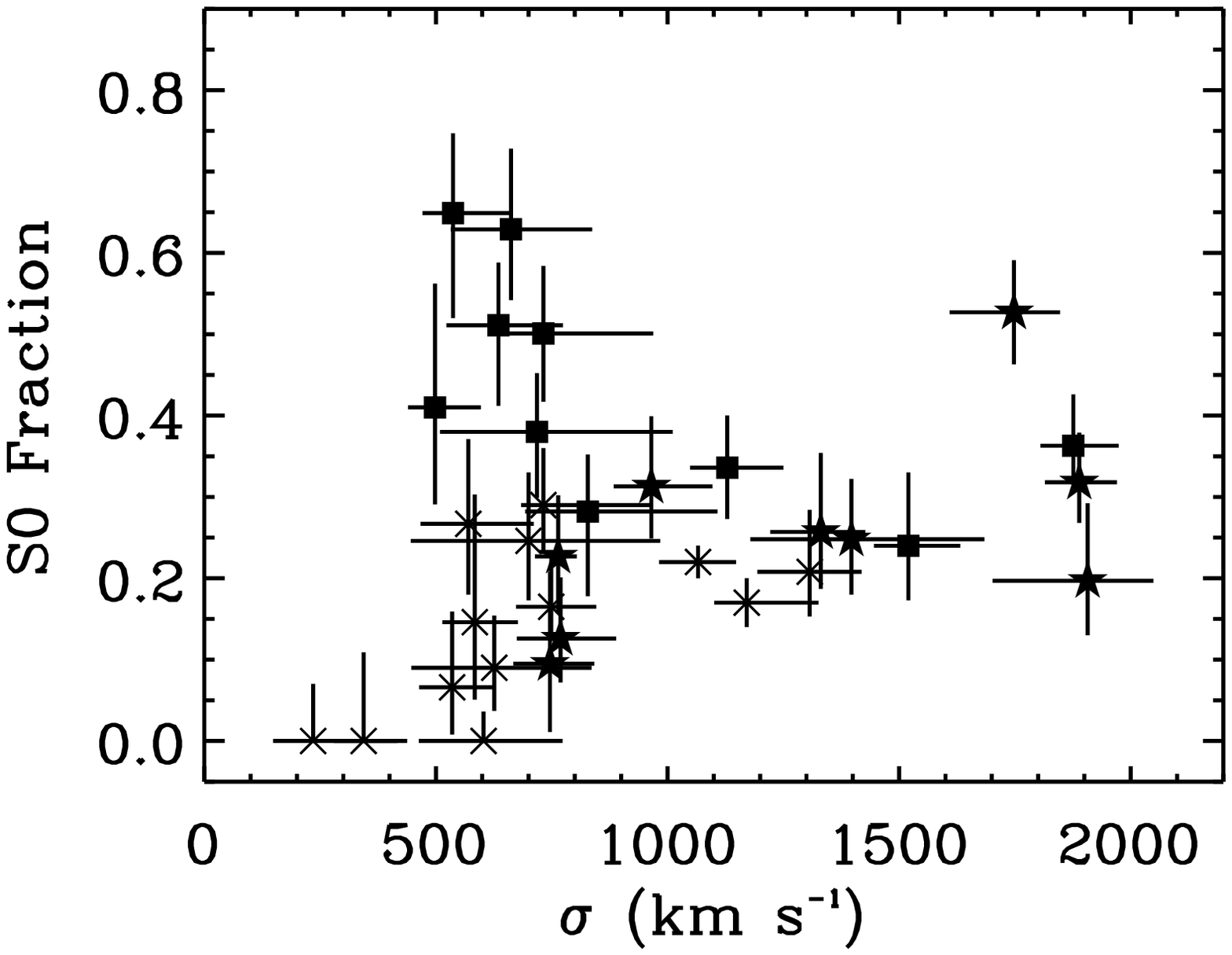} 
\figurenum{3}
\caption{\label{fs0vsig} S0 fraction (\fso) plotted against galaxy
cluster velocity dispersion ($\sigma$). There is no simple correlation
between these quantities but clearly a divergence of \fso\ at low
$\sigma$. The clusters with $z<0.3$ ({\it squares}) are entirely
non-MORPHS clusters from F00, the clusters from $0.3<z<0.5$ ({\it
stars}) are mostly MORPHS clusters from F00, and the clusters with
$z>0.5$ ({\it crosses}) are mostly EDisCS clusters from D07.}
\end{figure}

The results of the previous correlation analysis do not necessarily
imply that environment (velocity dispersion) plays no role. From
Figure~\ref{fs0vsig}, it is apparent that there is a wide spread in
\fso\ below $\sim 750$~km~s$^{-1}$ and a much narrower spread
above. We therefore split the sample into a high-$\sigma$ bin and a
low-$\sigma$ bin at this value to investigate the effect of
environment on the \fso-$z$ relation. This choice divides the sample
into nearly equal parts as well as into samples that are more typical
of groups/poor clusters (\losig) and rich clusters (\hisig). Some of
the clusters have suspiciously high velocity dispersions
($\sigma\gtsim1500 {\rm ~km~s}^{-1}$) and are presumably unrelaxed
systems (e.g., A1689). Nevertheless, given our gross binning scheme
they are still likely to be systems with \hisig\ and placed in the
appropriate velocity dispersion bin. Selecting a boundary anywhere up
to 1050~km~s$^{-1}$ or down to 650~km~s$^{-1}$ (after which the number
of clusters in the low-$\sigma$ bin drops sharply) leaves the results
that follow qualitatively unchanged, as does removing the clusters
with $\sigma\gtsim1500$~km~s$^{-1}$ from the analysis.

In Figure~\ref{highlow}, we show \fso\ plotted against redshift in the
high-$\sigma$ and low-$\sigma$ bins. While the \fso-$z$ trend is
evident in the groups/poor clusters, the correlation appears to be
much weaker, if present at all, in the rich clusters. Using
uncertainty-weighted least-squares fitting, we find that the slope for
the groups/poor clusters, $-0.75\pm0.14$, is steeper than the slope
for the rich clusters, $-0.18\pm0.09$ (a $3.4\sigma$ difference in
slope). For the high-$\sigma$ clusters, one may worry that there is
only one data point at $z>0.6$, which has an anomalously small error
of $\pm0.02$ and therefore strongly influences the slope. To explore
the impact of this one cluster on the fit, we have assigned it an
uncertainty equal to the scatter in \fso\ for the high-$\sigma$
clusters, $\pm 0.07$.  With this larger uncertainty estimate the new
slope is $-0.38\pm0.13$, resulting in only a $1.9\sigma$ difference in
slope between the low- and high-$\sigma$ clusters. To bolster the case
for the flat relationship among the massive clusters, we compare the
morphological fractions to those from \cite{Postman05}. Although we
argued in \S2.1 against using these clusters for our statistical
analyses, they support our finding that the relationship with redshift
is nearly flat for massive clusters (Figure~\ref{highlow}).  We
conclude that the difference in behavior between the low and
high-$\sigma$ clusters is not the result of the one high-z EDisCS
cluster. Lastly, the two lowest-$\sigma$ clusters in the EDisCS sample
have \fso~$=0$ and are potentially very unusual, although excluding
them from this analysis does not alter the results.

The clusters driving most of the trend in the groups/poor clusters are
the high-\fso\ systems at low $z$. Among those at $z<0.3$, there is an
apparent dichotomy between those with a dense concentration of
ellipticals toward the cluster center and those less centrally
concentrated, in the sense that the latter have higher \fso\
(F00). Therefore, it is also possible that S0 evolution depends
further on an environmental property marked by the distribution of
cluster ellipticals. Even so, there is an increase in \fso\ since
$z\sim~0.5$ (F00) when considering the high- and low-elliptical
concentration systems separately.

In Figure~\ref{fevz}, we show the elliptical fraction (\fe) plotted
against redshift for the entire sample, the low-$\sigma$ subsample,
and the high-$\sigma$ subsample. In all three cases, there is no
significant trend of \fe\ with redshift. This argues against a
misclassification between S0's and ellipticals as the origin of the S0
evolution.

\section{Discussion}\label{sec:discussion}

As we have described, previous studies have found a factor of $\sim3$
increase in \fso\ between $z\sim0.5$ and $z\sim0$, with a
corresponding decrease in the spiral fraction and a constant
elliptical fraction \citep{Dressler97,Fasano00}. Some authors
\citep[e.g.,][]{Andreon98} have noted that the trends, which at some
level must be affected by selection effects and methodology, may be a
result of unappreciated biases. The ability to distinguish between
S0's and ellipticals at higher redshifts, or other problems associated
with morphological classification, could in principle result in
spurious correlations. With this specific issue in mind, we
investigate the relationships of various morphological fractions with
redshift and velocity dispersion. We have already argued against a
redshift-dependent classification problem in E's vs S0's (see
above). What if there is an analogous problem with environment? For
example, if ellipticals are more common in the more massive
environments to the limits of our classification, and if a constant
fraction of those are misclassified as S0's, then $f_{S0}$ would
appear higher in more massive environments.  For Figure 5 we also
conclude that there is no discernible difference in the $f_E$ as a
function of environment over the range of environments explored here.

\begin{figure*}
\epsscale{1.0}
\plotone{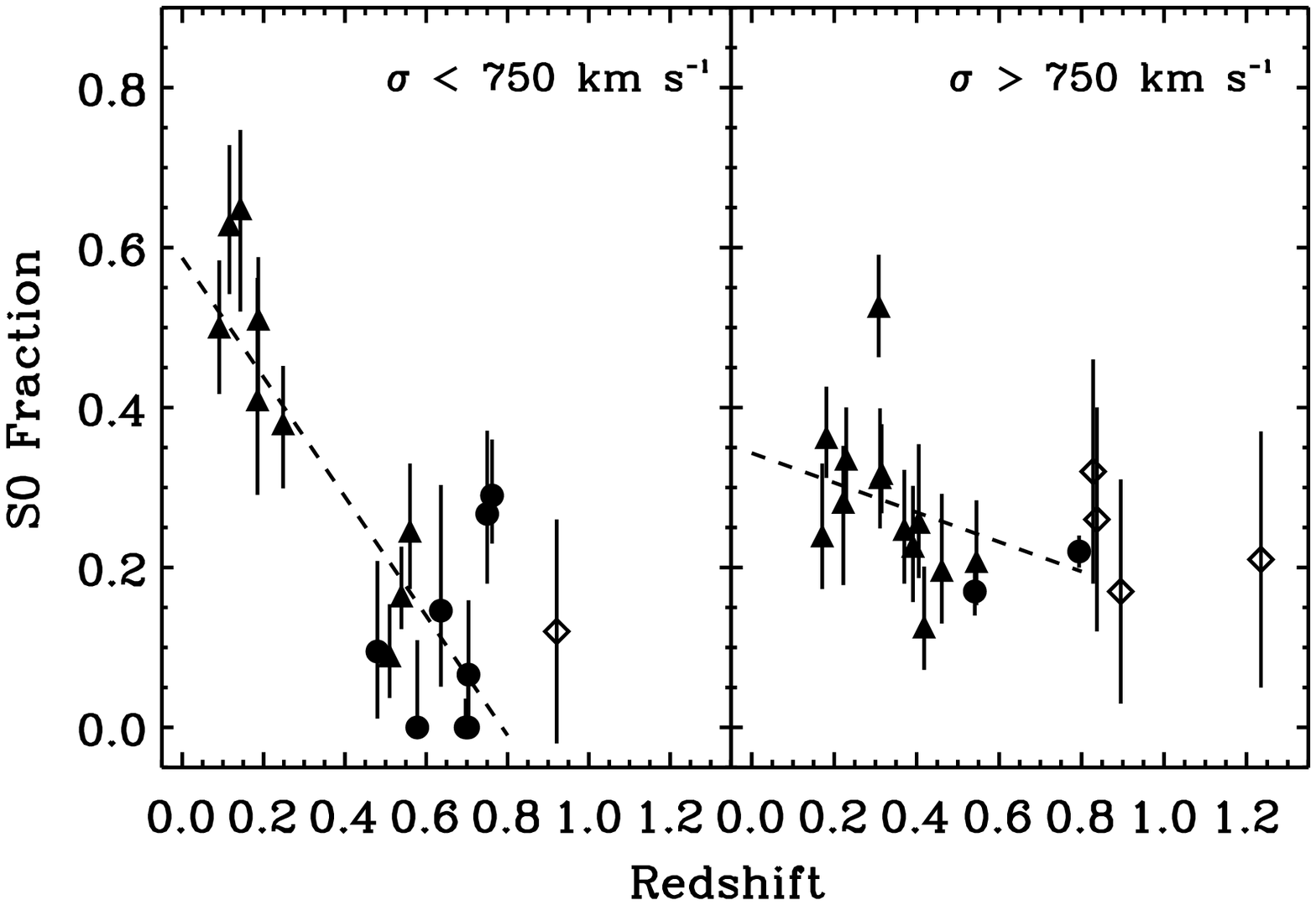}
\figurenum{4}
\caption{\label{highlow} S0 Fraction plotted against redshift in a
low-mass, $\sigma < 750$ km~s$^{-1}$ bin ({\it left}) and a high-mass,
$\sigma > 750$ km~s$^{-1}$ bin ({\it right}) for the F00 and EDisCS
clusters (triangles and circles, respectively); this binning roughly
splits the sample into groups/poor clusters and rich clusters,
respectively. The trend is clear in the groups/poor clusters sample
(with a slope of $-0.75\pm0.14$), but hardly evident in the rich
clusters (with a slope of $-0.18\pm0.09$), consistent with the idea
that morphological transformation is taking place in group/poor
cluster environments over this redshift range. The subset of clusters
from \cite{Postman05} with velocity dispersion measurements are
plotted as open diamonds; these clusters are {\it not} used in the
fits for reasons given in \S2.1 and are shown for illustrative
purposes only.}
\end{figure*}

We now remove the ellipticals from consideration and consider a plot
similar to Figure~\ref{highlow} in which we replace the ordinate,
\fso, with $N_{\rm S0}/(N_{\rm S}+N_{\rm S0})$, where $N_{\rm S0}$ and
$N_{\rm S}$ are the numbers of S0's and spirals in each cluster,
respectively (Figure~\ref{s0svz}). The dichotomy in the rate of
evolution between low-$\sigma$ groups/poor clusters and high-$\sigma$
rich clusters remains, with slopes of $-1.19\pm0.24$ and
$-0.07\pm0.17$, respectively (a 3.8$\sigma$ difference in slope). The
difference between the morphological fractions of the two environments
at low redshifts indicates that the morphological distinction between
spirals and S0's is reflecting a true underlying difference between
the two environments.  The difference in evolutionary trends does not,
unfortunately, necessarily imply that the trends are unaffected by
misclassification; if the two environments have different intrinsic
fractions of spirals and S0's, then redshift-dependent
misclassification could affect each environment differently.

Given the results described so far, we interpret \citep[as others
before have, e.g.][]{Dressler97,Fasano00,Smith05,Poggianti06} that the
evolving S0 fraction represents the transformation of spirals into
S0's. The difference here is that the S0 evolution (over these
redshifts) is taking place primarily in groups/poor clusters with
\losig\ (Figure~\ref{highlow}), suggesting that this is the location
of S0 formation. This result then supports the hypothesis that direct
galaxy interactions, i.e. mergers and/or close tidal encounters, are
the dominant mechanisms in converting spirals into S0's over the
redshift interval examined.  The value of $\sigma$ where galaxy-galaxy
processes dominate and where galaxy-environment process dominate is
not theoretically well constrained. Although we choose a cutoff at
$750$~km~s$^{-1}$ to divide the sample into equal parts, and expect
mergers and/or tidal interactions to dominate in the low-$\sigma$
subsample, the division into two subsamples only crudely reflects a
distinction of environments where different physical effects may
dominate.  However, the existence of high \fso\ systems with low
velocity dispersions demonstrates that {\it neither} the nature or
nurture of massive environments is necessary to the formation of S0's.

The conclusion that groups are the site of S0 formation, and therefore
that mergers/interactions are the formation mechanism, has been
arrived at in various ways. \cite{Wilman09} find a high \fso\ already
in place in $z \sim 0.5$ groups.  \cite{Poggianti09} find
more-pronounced S0 evolution in clusters with $\sigma\ltsim
800$~km~s$^{-1}$ by comparing a $z\sim 0$ sample to a high-z sample,
although their inclusion of the same EDisCS clusters means the results
are not entirely independent from ours. More distinctly,
\cite{Christlein04} find that S0's differ from normal spirals due to a
higher bulge luminosity rather than fainter disks, and interpret this
as requiring bulge growth during S0 formation. They conclude that such
formation mechanisms as strangulation and ram pressure stripping are
therefore disfavored. \cite{Hinz03} argue that the large scatter they
measure in the local S0 Tully-Fisher relation support formation
mechanisms that kinematically disturb the galaxies, i.e. interactions.
The unique aspect of our observations is that we establish both the
redshift and the environment at which this formation is
occurring. Thereby, we identify the exact place to focus further
investigation and perhaps distinguish the progenitors. Fortunately,
this evolution happens at redshifts that are relatively easily
accessed with current technology.

Although S0 evolution is seen primarily in the low-$\sigma$ clusters
and the values of \fso\ reach between 0.5 and 0.6 at $z \sim 0$, the
rate of S0 formation must reverse itself at some low value of the
velocity dispersion so as not to overpopulate the field with S0's
\citep[the local field \fso\ $\sim0.10$;][]{Sandage87}. Determining
this transitional value of the velocity dispersion would further aid
our understanding of the environmental processes at work. For example,
one might find that this velocity dispersion corresponds to that of
environments where the probability of interactions in a Hubble time
become unlikely (e.g. slightly more massive than the Local Group). Our
lowest-$z$ clusters extend down to $\sim500$ km~s$^{-1}$, while the
$z\sim0$ clusters of \cite{Poggianti09} probe down to $\sim400$
km~s$^{-1}$, setting an upper limit on where the trend must reverse
(our two lowest velocity dispersion systems, both with $\sigma < 400$
km s$^{-1}$, but high redshifts, have \fso\ $\sim 0$, perhaps
suggesting where this turnover occurs).

\begin{figure}
\epsscale{1.0}
\plotone{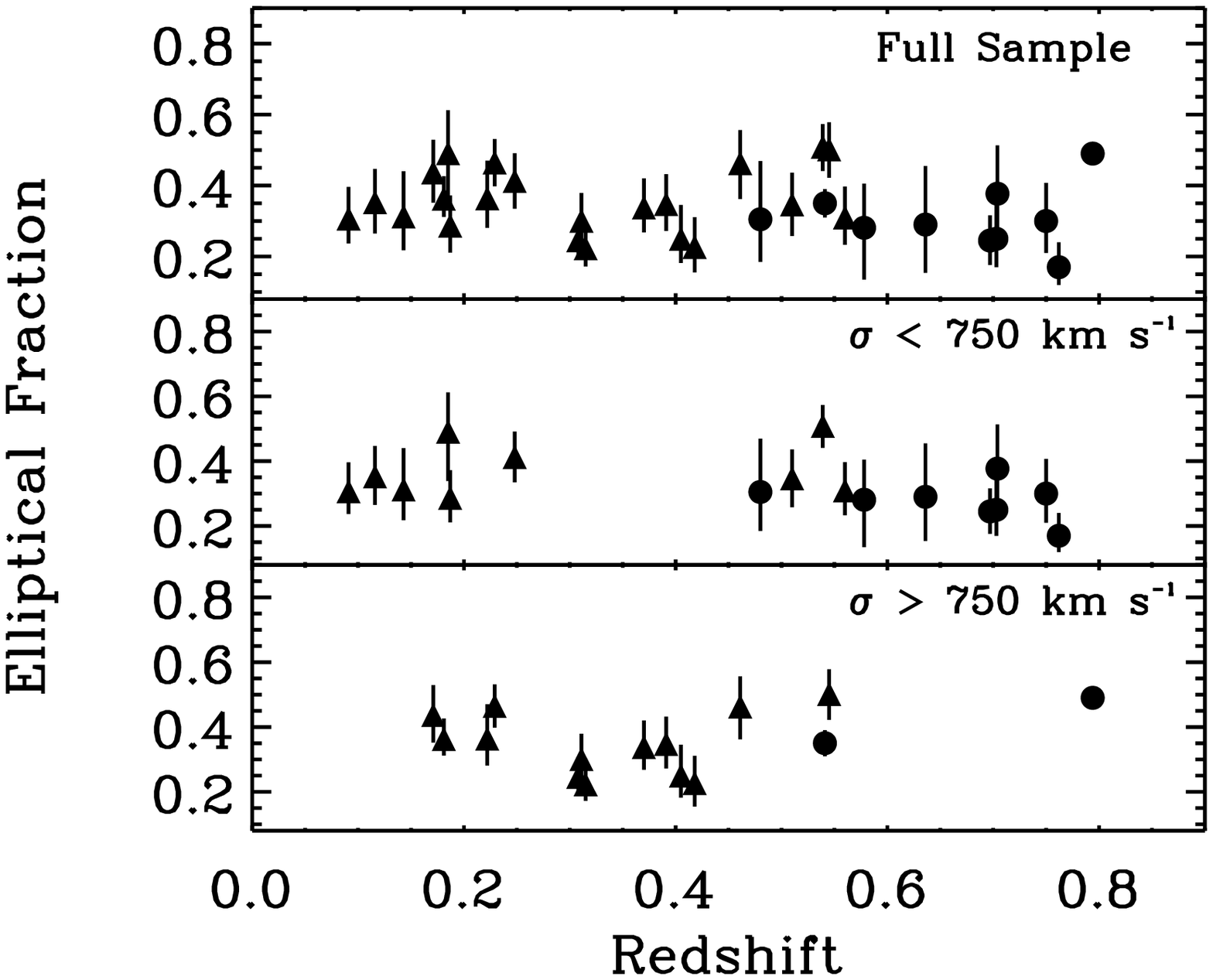}
\figurenum{5}
\caption{\label{fevz} Elliptical fraction ($f_{\rm E}$) plotted
against redshift for the full sample ({\it top}), the low-mass,
$\sigma < 750$ km~s$^{-1}$ bin ({\it middle}), and the high-mass,
$\sigma > 750$ km~s$^{-1}$ bin ({\it bottom}). Symbols are the same as
in Figure 2. Neither the full sample nor the subsamples show a
significant trend in elliptical fraction with redshift.}
\end{figure}

So far, we have not accounted for the effects of the hierarchical
growth of groups and clusters on the question of S0 evolution. Groups
and clusters grow over time, accreting galaxies from the field and/or
groups, so that systems at $z\sim0.8$ with a particular value of
$\sigma$ do not correspond to those of the same $\sigma$ at $z=0$. It
has generally been assumed, due to the expectation that S0's would be
rarer in low density environments, that any accretion these systems
experience would be S0-poor, hence the need to transform some fraction
of these galaxies into S0's.  From Figure~{\ref{highlow}}, we now know
that this is not the case, at least for $z < 0.3$. In fact, at low $z$
it appears that high-$z$ clusters could increase their \fso\ over time
by accreting these smaller systems without requiring any morphological
transformation mechanism. How much of the observed \fso-$z$ trend in
the high-$\sigma$ rich clusters could simply be due to the accretion
of smaller, S0-rich groups/poor clusters similar to those in our
low-$\sigma$ subsample?

To estimate the increase in the number of cluster galaxies with
redshift, we note that the mass of rich clusters at $z\sim0.5$
typically increases $\sim40\%$ by $z=0$ \citep{Wechsler02}, and assume
that this increase in mass corresponds to the same relative increase
in the number of cluster galaxies. We also assume that the mass
accretion comes in the form of our low-$\sigma$ groups.  To the degree
that field galaxies, with their lower $f_{S0}$, account for the
accreted mass then this model will be an overestimate of the effect.
The final S0 fraction $f_{{\rm S0},z=0}$ in this simple model is
\begin{equation}
f_{{\rm S0},z=0}=\frac{f_{{\rm S0},z=0.5}+\eta f_{\rm S0,gr}}{1+\eta},
\end{equation}
where $f_{{\rm S0},z=0.5}$ is the S0 fraction of the cluster at
$z=0.5$, $\eta$ is the fractional increase in number of cluster
galaxies from $z=0.5$ to $z=0$, i.e. $\eta=0.4$ based on the
\cite{Wechsler02} models, and $f_{\rm S0,gr}$ is the S0 fraction for
low-$z$ groups, for which we adopt a conservative value of 0.4. From
our best-fit trend in the high-$\sigma$ panel of
Figure~{\ref{highlow}}, the S0 fraction for a massive cluster at $z =
0.5$, $f_{{\rm S0},z=0.5}$, is 0.25. Using Equation~(4) gives $f_{{\rm
S0},z=0}\approx0.3$, consistent with our best-fit trend at
$z=0$. Therefore, this simple model suggests that the trend of
increasing \fso\ with $z$ in the high-$\sigma$ clusters {\sl could} be
accounted for solely by the accretion of S0-rich groups. Regardless of
the actual accretion history, we conclude that the accretion of at
least some S0-rich groups will explain part of the increase in \fso\
in clusters.

\begin{figure}
\epsscale{1.0}
\plotone{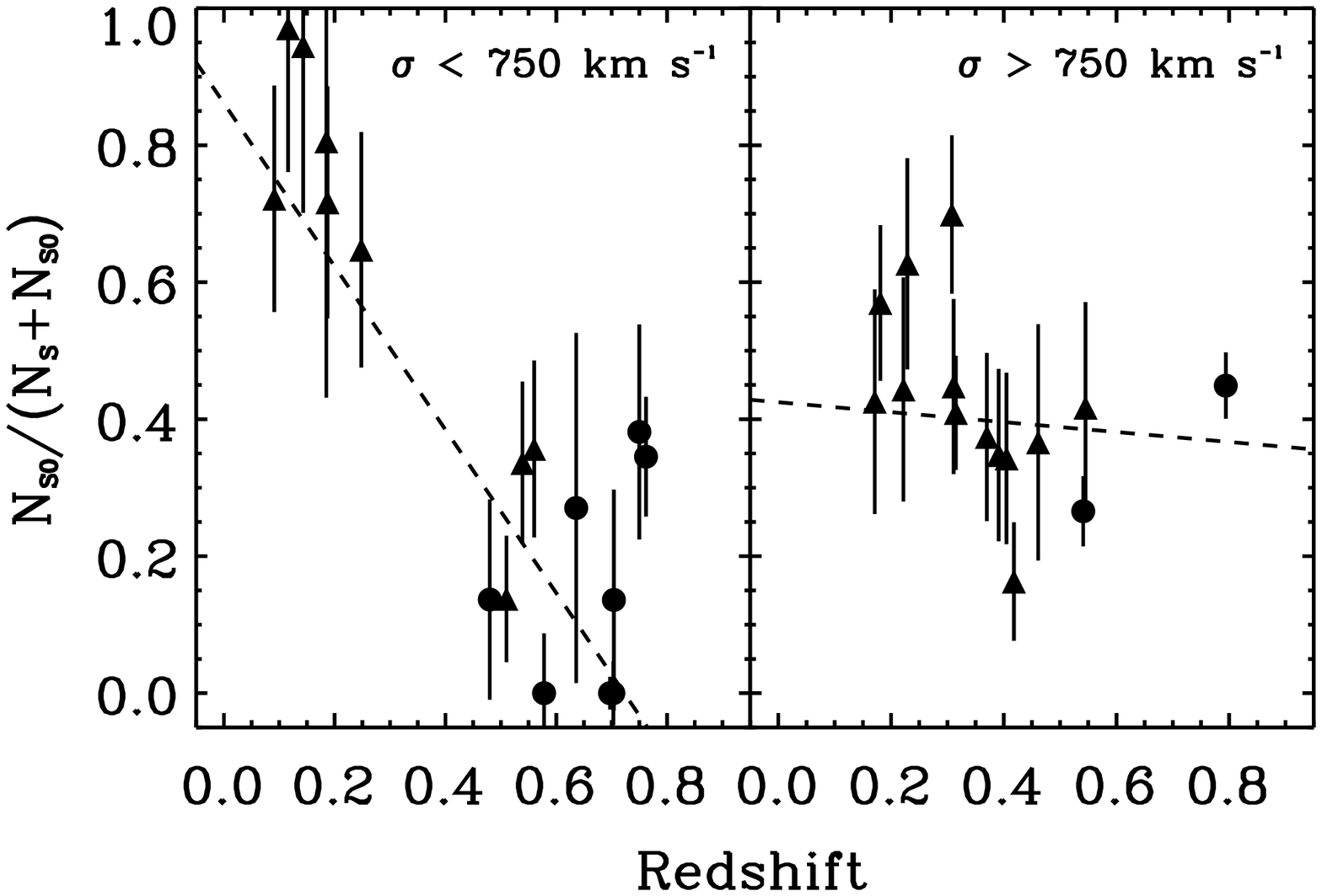}
\figurenum{6}
\caption{\label{s0svz} $N_{\rm S0}/(N_{\rm S}+N_{\rm S0})$, where
$N_{\rm S0}$ and $N_{\rm S}$ are the numbers of S0's and spirals,
respectively, plotted against redshift in the low-mass, $\sigma <
750$~km~s$^{-1}$ bin ({\it left}) and the high-mass, $\sigma >
750$~km~s$^{-1}$ bin ({\it right}). The dashed line shows best-fit
trends, with significantly different slopes of $-1.19\pm0.24$ and
$-0.07\pm0.17$ (a 3.8$\sigma$ difference in slope) in the left and
right panels, respectively. Symbols are the same as in Figure 2.}
\end{figure}

The results presented here (and elsewhere) that S0 galaxies are
forming at relatively low redshifts ($z < 0.5$) and in low-$\sigma$
groups, implies that we should be able to identify and study both the
progenitor class and the galaxies undergoing this transition.
Post-starburst galaxies are commonly suspected to be late-time
examples of the latter \citep{Dressler85,Couch87,Yang04,Yang06}. If
so, this transformation affects both the morphology and stellar
population of the galaxy and we expect based on our results that 1)
S0's in rich clusters at $z=0$ will have mostly old stellar
populations ($\gtsim7$~Gyr) because most of their S0 population has
been in place since $z\sim0.8$ and 2) the S0's in low-$\sigma$, $z=0$
clusters will have a mix of young and old stars, with roughly 50\% of
the S0's having a significant fraction of their stars that are younger
than $\sim3$~Gyr old \citep[evidence for some relatively young S0
galaxies in the field now exists;][]{Moran07,Kannapann09}.

\section{Conclusion}\label{sec:conclusion}

By compiling a large set of clusters with both internally-consistent
morphological classifications and uniform velocity dispersions,
$\sigma$, we examined the rate of change in the S0 fraction, \fso,
with redshift as a function of environment. We show that for our
entire sample \fso\ is primarily correlated with redshift and not
significantly correlated with velocity dispersion. However, the
evolution of \fso\ with redshift is much stronger among
$\sigma<750$~km~s$^{-1}$ galaxy groups/poor clusters than in
higher-$\sigma$ rich clusters. We interpret this result to mean that
direct processes like galaxy mergers, which are expected to dominate
in lower-$\sigma$ environments, are the primary mechanisms for
morphological transformation over the redshift range explored, $0 < z
\ltsim 0.8$.

Further studies would benefit from a larger sample size, in particular
having \fso\ and $\sigma$ measurements for both groups/poor clusters
and rich clusters with comparable numbers across a similar range in
redshift. This study highlights the importance of having velocity
dispersion measurements in evolutionary studies, so that one can
account for any environmental dependence of the evolution itself. In
particular, we emphasize that more complete samples of environments
are needed and that large numbers of redshifts per system are
necessary to convincingly measure velocity dispersions of low-mass
systems. Lastly, as emphasized by \cite{Dressler80} and
\cite{Postman84}, local density may be a critical factor in S0
formation. We cannot measure the evolution of \fso\ as a function of
local density from our data due to the small number of spectroscopic
members per system, but both larger cluster/group samples and more
redshifts per system would enable such a study.

\acknowledgements We thank the anonymous referee for insightful
comments that improved the content and presentation of his paper. DJ
thanks Michael Cooper for useful conversations and Daniel Christlein
for helping with the Magellan observations. DZ acknowledges financial
support for this work from NASA LTSA award NNG05GE82G, NASA XMM grants
NNX06AG39A, NNX06AE41G, and NASA Spitzer grant 1344985. DZ also thanks
the NYU Physics Department and CCPP for their hospitality during his
visit.

\begin{figure*}
\epsscale{0.8}
\plotone{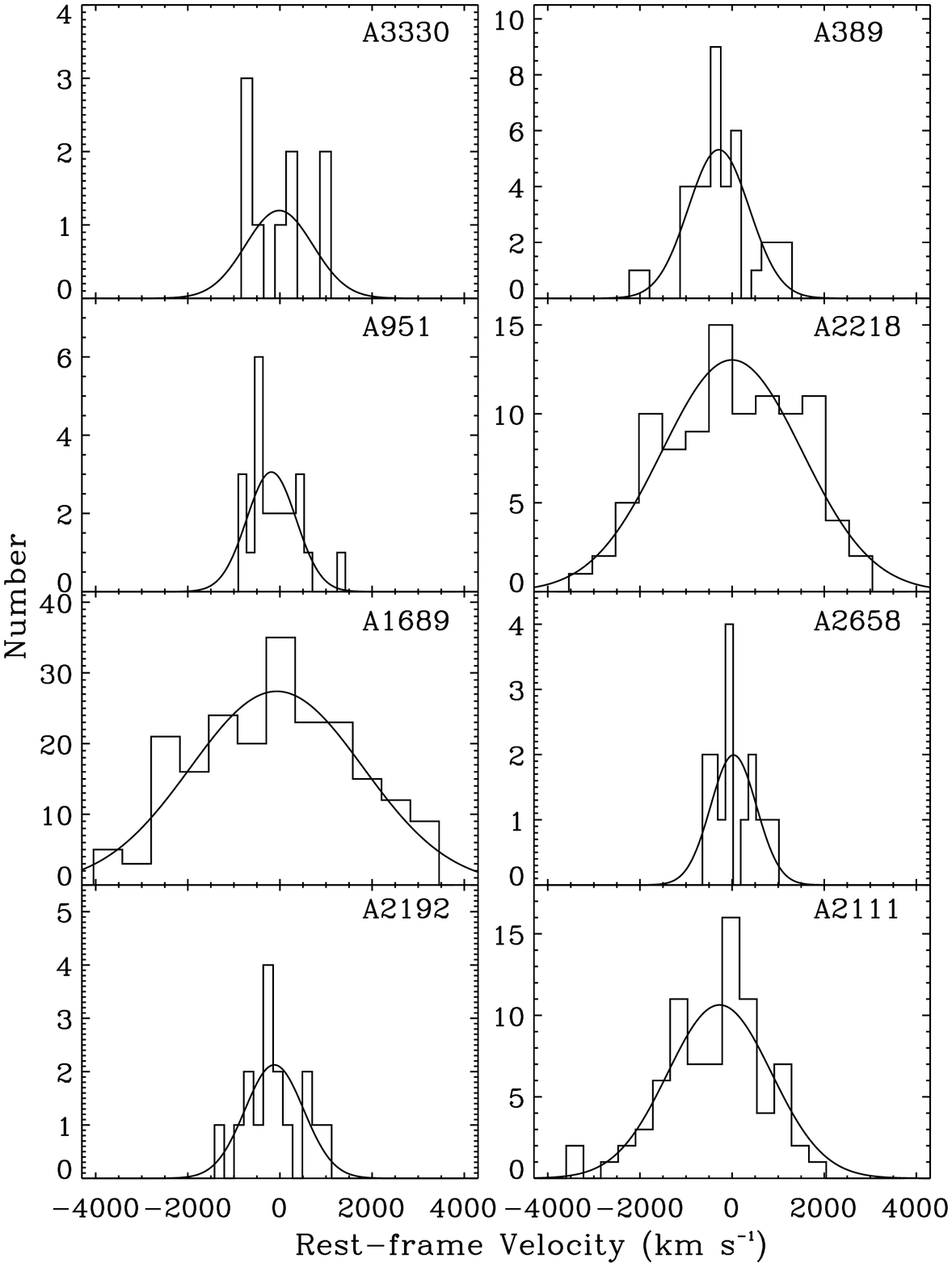}
\figurenum{1}
\caption{\label{hist} Rest-frame velocity histograms. The bin size is
one-third the velocity dispersion, and the velocities plotted are
those that remain after the various cuts/iterations in the calculation
of $\sigma$. Overplotted on each panel is a Gaussian normalized to the
area of the histogram.}
\end{figure*}

\begin{figure*}
\epsscale{0.8}
\plotone{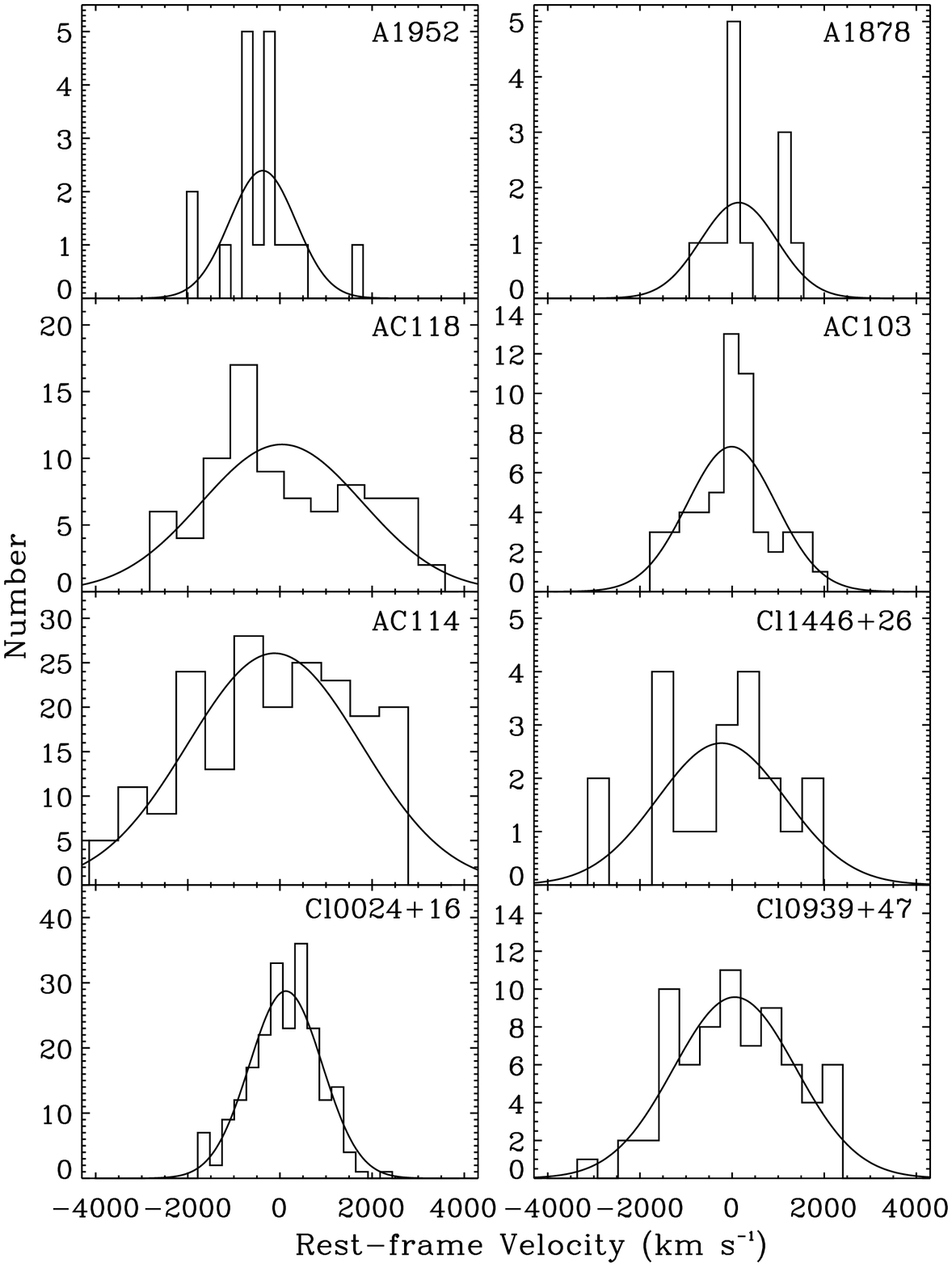}
\figurenum{1}
\caption{{\it Continued}}
\end{figure*}

\begin{figure*}
\epsscale{0.8}
\plotone{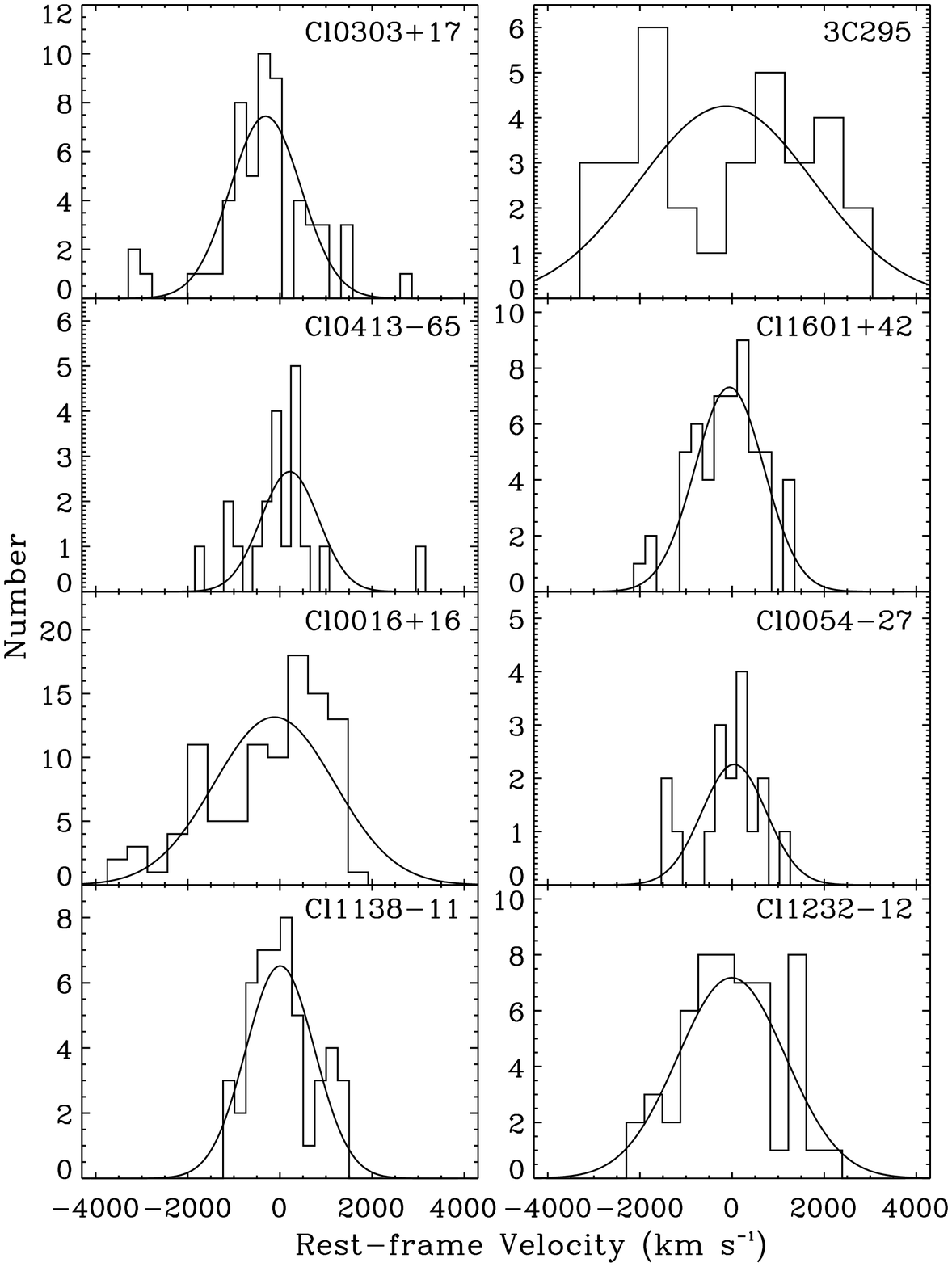}
\figurenum{1}
\caption{{\it Continued}}
\end{figure*}

\begin{figure*}
\epsscale{0.8}
\plotone{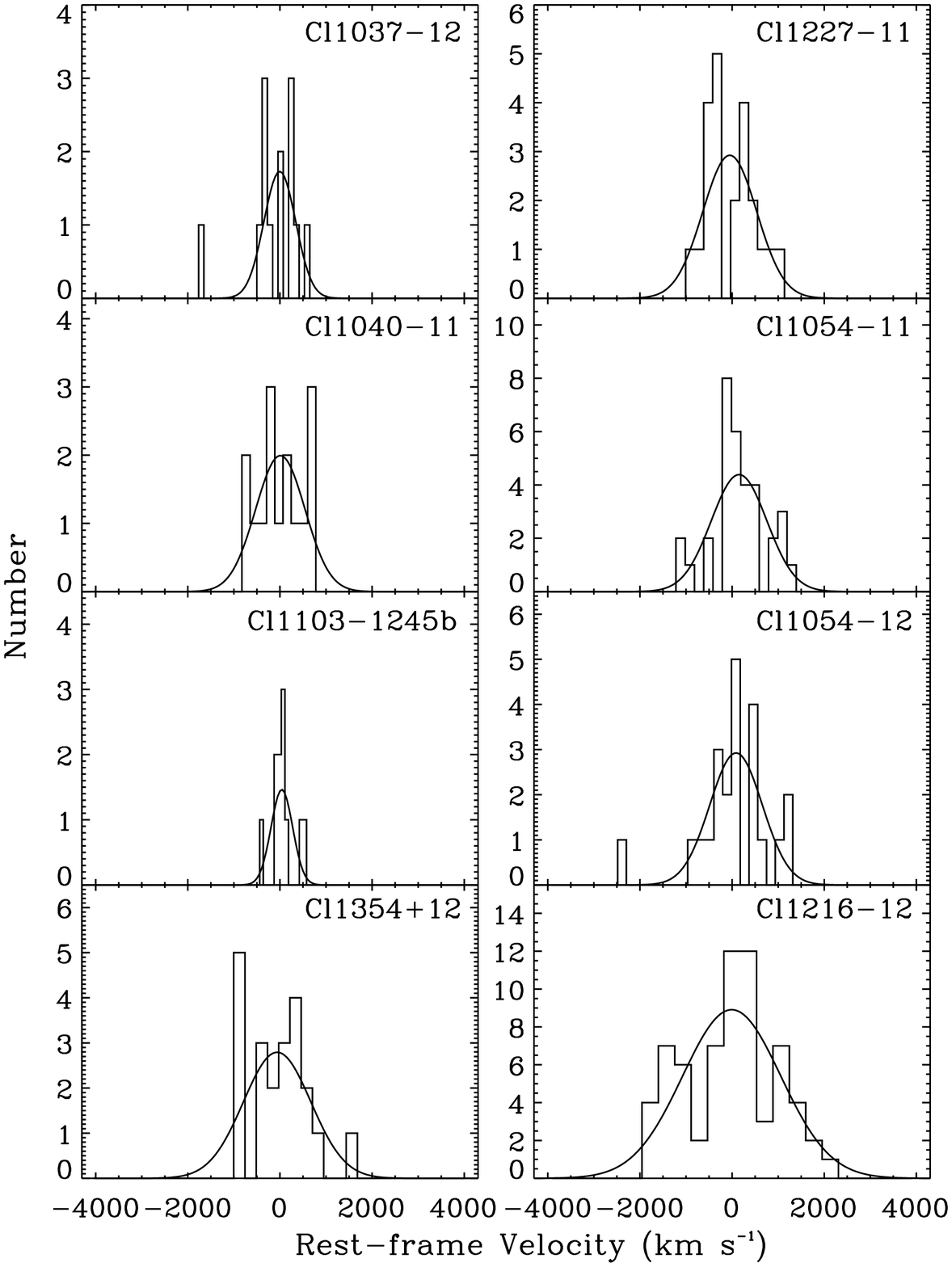}
\figurenum{1}
\caption{{\it Continued}}
\end{figure*}


\begin{thebibliography}
\expandafter\ifx\csname natexlab\endcsname\relax\def\natexlab#1{#1}\fi

\bibitem[{{Abadi} {et~al.}(1999){Abadi}, {Moore}, \& {Bower}}]{Abadi99}
{Abadi}, M.~G., {Moore}, B., \& {Bower}, R.~G. 1999, \mnras, 308, 947

\bibitem[{{Andreon}(1998)}]{Andreon98}
{Andreon}, S. 1998, \apj, 501, 533

\bibitem[{{Balogh} {et~al.}(2004){Balogh}, {Baldry}, {Nichol}, {Miller},
  {Bower}, \& {Glazebrook}}]{Balogh04}
{Balogh}, M.~L., {Baldry}, I.~K., {Nichol}, R., {Miller}, C., {Bower}, R., \&
  {Glazebrook}, K. 2004, \apjl, 615, L101

\bibitem[{{Balogh} {et~al.}(2002){Balogh}, {Smail}, {Bower}, {Ziegler},
  {Smith}, {Davies}, {Gaztelu}, {Kneib}, \& {Ebeling}}]{Balogh02}
{Balogh}, M.~L., {et~al.} 2002, \apj, 566, 123

\bibitem[{{Beers} {et~al.}(1990){Beers}, {Flynn}, \& {Gebhardt}}]{Beers90}
{Beers}, T.~C., {Flynn}, K., \& {Gebhardt}, K. 1990, \aj, 100, 32

\bibitem[{{Bekki} {et~al.}(2002){Bekki}, {Couch}, \& {Shioya}}]{Bekki02}
{Bekki}, K., {Couch}, W.~J., \& {Shioya}, Y. 2002, \apj, 577, 651

\bibitem[{{Bigelow} {et~al.}(1998){Bigelow}, {Dressler}, {Shectman}, \&
  {Epps}}]{Bigelow98}
{Bigelow}, B.~C., {Dressler}, A.~M., {Shectman}, S.~A., \& {Epps}, H.~W. 1998,
  in Presented at the Society of Photo-Optical Instrumentation Engineers (SPIE)
  Conference, Vol. 3355, Society of Photo-Optical Instrumentation Engineers
  (SPIE) Conference Series, ed. S.~{D'Odorico}, 225--231

\bibitem[{{Byrd} \& {Valtonen}(1990)}]{Byrd90}
{Byrd}, G., \& {Valtonen}, M. 1990, \apj, 350, 89

\bibitem[{{Christlein} \& {Zabludoff}(2004)}]{Christlein04}
{Christlein}, D., \& {Zabludoff}, A.~I. 2004, \apj, 616, 192

\bibitem[{{Couch} {et~al.}(1998){Couch}, {Barger}, {Smail}, {Ellis}, \&
  {Sharples}}]{Couch98}
{Couch}, W.~J., {Barger}, A.~J., {Smail}, I., {Ellis}, R.~S., \& {Sharples},
  R.~M. 1998, \apj, 497, 188

\bibitem[{{Couch} \& {Sharples}(1987)}]{Couch87}
{Couch}, W.~J., \& {Sharples}, R.~M. 1987, \mnras, 229, 423

\bibitem[{{Desai} {et~al.}(2007){Desai}, {Dalcanton}, {Arag{\'o}n-Salamanca},
  {Jablonka}, {Poggianti}, {Gogarten}, {Simard}, {Milvang-Jensen}, {Rudnick},
  {Zaritsky}, {Clowe}, {Halliday}, {Pell{\'o}}, {Saglia}, \& {White}}]{Desai07}
{Desai}, V., {et~al.} 2007, \apj, 660, 1151

\bibitem[{{Dressler}(1980)}]{Dressler80}
{Dressler}, A. 1980, \apj, 236, 351

\bibitem[{{Dressler} {et~al.}(1985){Dressler}, {Gunn}, \&
  {Schneider}}]{Dressler85}
{Dressler}, A., {Gunn}, J.~E., \& {Schneider}, D.~P. 1985, \apj, 294, 70

\bibitem[{{Dressler} {et~al.}(1997){Dressler}, {Oemler}, {Couch}, {Smail},
  {Ellis}, {Barger}, {Butcher}, {Poggianti}, \& {Sharples}}]{Dressler97}
{Dressler}, A., {et~al.} 1997, \apj, 490, 577

\bibitem[{{Fabricant} {et~al.}(2005){Fabricant}, {Fata}, {Roll}, {Hertz},
  {Caldwell}, {Gauron}, {Geary}, {McLeod}, {Szentgyorgyi}, {Zajac}, {Kurtz},
  {Barberis}, {Bergner}, {Brown}, {Conroy}, {Eng}, {Geller}, {Goddard},
  {Honsa}, {Mueller}, {Mink}, {Ordway}, {Tokarz}, {Woods}, {Wyatt}, {Epps}, \&
  {Dell'Antonio}}]{Fabricant05}
{Fabricant}, D., {et~al.} 2005, \pasp, 117, 1411

\bibitem[{{Fasano} {et~al.}(2000){Fasano}, {Poggianti}, {Couch}, {Bettoni},
  {Kj{\ae}rgaard}, \& {Moles}}]{Fasano00}
{Fasano}, G., {Poggianti}, B.~M., {Couch}, W.~J., {Bettoni}, D.,
  {Kj{\ae}rgaard}, P., \& {Moles}, M. 2000, \apj, 542, 673

\bibitem[{{Finn} {et~al.}(2008){Finn}, {Balogh}, {Zaritsky}, {Miller}, \&
  {Nichol}}]{Finn08}
{Finn}, R.~A., {Balogh}, M.~L., {Zaritsky}, D., {Miller}, C.~J., \& {Nichol},
  R.~C. 2008, \apj, 679, 279

\bibitem[{{Finn} {et~al.}(2004){Finn}, {Zaritsky}, \& {McCarthy}}]{Finn04}
{Finn}, R.~A., {Zaritsky}, D., \& {McCarthy}, Jr., D.~W. 2004, \apj, 604, 141

\bibitem[{{Finn} {et~al.}(2009){Finn}, {Desai}, {Rudnick}, {Lai}, {Hinz},
  {Cowen}, {Moustakas}, {Jablonka}, {Zaritsky}, {Milvang-Jensen}, {Bell},
  {McIntosh}, \& {Rines}}]{Finn09}
{Finn}, R.~A., {et~al.} 2009, in prep

\bibitem[{{Gehrels}(1986)}]{Gehrels86}
{Gehrels}, N. 1986, \apj, 303, 336

\bibitem[{{Gunn} \& {Gott}(1972)}]{Gunn72}
{Gunn}, J.~E., \& {Gott}, J.~R.~I. 1972, \apj, 176, 1

\bibitem[{{Halliday} {et~al.}(2004){Halliday}, {Milvang-Jensen}, {Poirier},
  {Poggianti}, {Jablonka}, {Arag{\'o}n-Salamanca}, {Saglia}, {De Lucia},
  {Pell{\'o}}, {Simard}, {Clowe}, {Rudnick}, {Dalcanton}, {White}, \&
  {Zaritsky}}]{Halliday04}
{Halliday}, C., {et~al.} 2004, \aap, 427, 397

\bibitem[{{Helsdon} \& {Ponman}(2003)}]{Helsdon03}
{Helsdon}, S.~F., \& {Ponman}, T.~J. 2003, \mnras, 339, L29

\bibitem[{{Hinz} {et~al.}(2003){Hinz}, {Rieke}, \& {Caldwell}}]{Hinz03}
{Hinz}, J.~L., {Rieke}, G.~H., \& {Caldwell}, N. 2003, \aj, 126, 2622

\bibitem[{{Holden} {et~al.}(2009){Holden}, {Franx}, {Illingworth}, {Postman},
  {van der Wel}, {Kelson}, {Blakeslee}, {Ford}, {Demarco}, \& {Mei}}]{Holden09}
{Holden}, B.~P., {et~al.} 2009, \apj, 693, 617

\bibitem[{{Icke}(1985)}]{Icke85}
{Icke}, V. 1985, \aap, 144, 115

\bibitem[{{Kannappan} {et~al.}(2009){Kannappan}, {Guie}, \&
  {Baker}}]{Kannapann09}
{Kannappan}, S.~J., {Guie}, J.~M., \& {Baker}, A.~J. 2009, \aj, 138, 579

\bibitem[{{Kautsch} {et~al.}(2008){Kautsch}, {Gonzalez}, {Soto}, {Tran},
  {Zaritsky}, \& {Moustakas}}]{Kautsch08}
{Kautsch}, S.~J., {Gonzalez}, A.~H., {Soto}, C.~A., {Tran}, K.-V.~H.,
  {Zaritsky}, D., \& {Moustakas}, J. 2008, \apjl, 688, L5

\bibitem[{{Kawata} \& {Mulchaey}(2008)}]{Kawata08}
{Kawata}, D., \& {Mulchaey}, J.~S. 2008, \apjl, 672, L103

\bibitem[{{Kendall} \& {Stuart}(1977)}]{Kendall77}
{Kendall}, M., \& {Stuart}, A. 1977, {The advanced theory of statistics. Vol.1:
  Distribution theory}

\bibitem[{{Larson} {et~al.}(1980){Larson}, {Tinsley}, \& {Caldwell}}]{Larson80}
{Larson}, R.~B., {Tinsley}, B.~M., \& {Caldwell}, C.~N. 1980, \apj, 237, 692

\bibitem[{{Lavery} \& {Henry}(1988)}]{Lavery88}
{Lavery}, R.~J., \& {Henry}, J.~P. 1988, \apj, 330, 596

\bibitem[{{Mihos}(2004)}]{Mihos04}
{Mihos}, J.~C. 2004, in Clusters of Galaxies: Probes of Cosmological Structure
  and Galaxy Evolution, ed. J.~S. {Mulchaey}, A.~{Dressler}, \& A.~{Oemler},
  277--+

\bibitem[{{Milvang-Jensen} {et~al.}(2008){Milvang-Jensen}, {Noll}, {Halliday},
  {Poggianti}, {Jablonka}, {Arag{\'o}n-Salamanca}, {Saglia}, {Nowak}, {von der
  Linden}, {De Lucia}, {Pell{\'o}}, {Moustakas}, {Poirier}, {Bamford}, {Clowe},
  {Dalcanton}, {Rudnick}, {Simard}, {White}, \& {Zaritsky}}]{MilvangJensen08}
{Milvang-Jensen}, B., {et~al.} 2008, \aap, 482, 419

\bibitem[{{Moore} {et~al.}(1998){Moore}, {Lake}, \& {Katz}}]{Moore98}
{Moore}, B., {Lake}, G., \& {Katz}, N. 1998, \apj, 495, 139

\bibitem[{{Moran} {et~al.}(2007){Moran}, {Ellis}, {Treu}, {Smith}, {Rich}, \&
  {Smail}}]{Moran07}
{Moran}, S.~M., {Ellis}, R.~S., {Treu}, T., {Smith}, G.~P., {Rich}, R.~M., \&
  {Smail}, I. 2007, \apj, 671, 1503

\bibitem[{{Papovich} {et~al.}(2006){Papovich}, {Cool}, {Eisenstein}, {Le
  Floc'h}, {Fan}, {Kennicutt}, {Smith}, {Rieke}, \& {Vestergaard}}]{Papovich06}
{Papovich}, C., {et~al.} 2006, \aj, 132, 231

\bibitem[{{Pello}{~et~al.}(2009)}]{pello}
{Pello}, R. {~et~al.} 2009, \aa, in press

\bibitem[{{Poggianti} {et~al.}(1999){Poggianti}, {Smail}, {Dressler}, {Couch},
  {Barger}, {Butcher}, {Ellis}, \& {Oemler}}]{Poggianti99}
{Poggianti}, B.~M., {Smail}, I., {Dressler}, A., {Couch}, W.~J., {Barger},
  A.~J., {Butcher}, H., {Ellis}, R.~S., \& {Oemler}, A.~J. 1999, \apj, 518, 576

\bibitem[{{Poggianti} {et~al.}(2006){Poggianti}, {von der Linden}, {De Lucia},
  {Desai}, {Simard}, {Halliday}, {Arag{\'o}n-Salamanca}, {Bower}, {Varela},
  {Best}, {Clowe}, {Dalcanton}, {Jablonka}, {Milvang-Jensen}, {Pello},
  {Rudnick}, {Saglia}, {White}, \& {Zaritsky}}]{Poggianti06}
{Poggianti}, B.~M., {et~al.} 2006, \apj, 642, 188

\bibitem[{{Poggianti} {et~al.}(2009){Poggianti}, {Fasano}, {Bettoni}, {Cava},
  {Dressler}, {Vanzella}, {Varela}, {Couch}, {D'Onofrio}, {Fritz},
  {Kjaergaard}, {Moles}, \& {Valentinuzzi}}]{Poggianti09}
---. 2009, \apjl, 697, L137

\bibitem[{{Postman} \& {Geller}(1984)}]{Postman84}
{Postman}, M., \& {Geller}, M.~J. 1984, \apj, 281, 95

\bibitem[{{Postman} {et~al.}(2005){Postman}, {Franx}, {Cross}, {Holden},
  {Ford}, {Illingworth}, {Goto}, {Demarco}, {Rosati}, {Blakeslee}, {Tran},
  {Ben{\'{\i}}tez}, {Clampin}, {Hartig}, {Homeier}, {Ardila}, {Bartko},
  {Bouwens}, {Bradley}, {Broadhurst}, {Brown}, {Burrows}, {Cheng}, {Feldman},
  {Golimowski}, {Gronwall}, {Infante}, {Kimble}, {Krist}, {Lesser}, {Martel},
  {Mei}, {Menanteau}, {Meurer}, {Miley}, {Motta}, {Sirianni}, {Sparks}, {Tran},
  {Tsvetanov}, {White}, \& {Zheng}}]{Postman05}
{Postman}, M., {et~al.} 2005, \apj, 623, 721

\bibitem[{{Quilis} {et~al.}(2000){Quilis}, {Moore}, \& {Bower}}]{Quilis00}
{Quilis}, V., {Moore}, B., \& {Bower}, R. 2000, Science, 288, 1617

\bibitem[{{Richstone}(1976)}]{Richstone76}
{Richstone}, D.~O. 1976, \apj, 204, 642

\bibitem[Sandage et al.(1976)]{Sandage76} Sandage, A., Kristian, J.,
\& Westphal, J.~A.\ 1976, \apj, 205, 688

\bibitem[{{Sandage} \& {Tammann}(1987)}]{Sandage87}
{Sandage}, A., \& {Tammann}, G.~A. 1987, {A revised Shapley-Ames Catalog of
  bright galaxies}

\bibitem[{{Smail} {et~al.}(1997){Smail}, {Dressler}, {Couch}, {Ellis},
  {Oemler}, {Butcher}, \& {Sharples}}]{Smail97}
{Smail}, I., {Dressler}, A., {Couch}, W.~J., {Ellis}, R.~S., {Oemler}, A.~J.,
  {Butcher}, H., \& {Sharples}, R.~M. 1997, \apjs, 110, 213

\bibitem[{{Smith} {et~al.}(2005){Smith}, {Treu}, {Ellis}, {Moran}, \&
  {Dressler}}]{Smith05}
{Smith}, G.~P., {Treu}, T., {Ellis}, R.~S., {Moran}, S.~M., \& {Dressler}, A.
  2005, \apj, 620, 78

\bibitem[{{Toomre} \& {Toomre}(1972)}]{Toomre72}
{Toomre}, A., \& {Toomre}, J. 1972, \apj, 178, 623

\bibitem[{{Tran} {et~al.}(2009){Tran}}]{Tran09}
{Tran, K.-V. H.,}{et~al.} 2009, \apj, 705, 809

\bibitem[{{Wechsler} {et~al.}(2002){Wechsler}, {Bullock}, {Primack},
  {Kravtsov}, \& {Dekel}}]{Wechsler02}
{Wechsler}, R.~H., {Bullock}, J.~S., {Primack}, J.~R., {Kravtsov}, A.~V., \&
  {Dekel}, A. 2002, \apj, 568, 52

\bibitem[{{White} {et~al.}(2005){White}, {Clowe}, {Simard}, {Rudnick}, {De
  Lucia}, {Arag{\'o}n-Salamanca}, {Bender}, {Best}, {Bremer}, {Charlot},
  {Dalcanton}, {Dantel}, {Desai}, {Fort}, {Halliday}, {Jablonka}, {Kauffmann},
  {Mellier}, {Milvang-Jensen}, {Pell{\'o}}, {Poggianti}, {Poirier},
  {Rottgering}, {Saglia}, {Schneider}, \& {Zaritsky}}]{White05}
{White}, S.~D.~M., {et~al.} 2005, \aap, 444, 365

\bibitem[{{Wilman} {et~al.}(2009){Wilman}, {Oemler}, {Mulchaey}, {McGee},
  {Balogh}, \& {Bower}}]{Wilman09}
{Wilman}, D.~J., {Oemler}, A., {Mulchaey}, J.~S., {McGee}, S.~L., {Balogh},
  M.~L., \& {Bower}, R.~G. 2009, \apj, 692, 298

\bibitem[{{Yang} {et~al.}(2006){Yang}, {Tremonti}, {Zabludoff}, \&
  {Zaritsky}}]{Yang06}
{Yang}, Y., {Tremonti}, C.~A., {Zabludoff}, A.~I., \& {Zaritsky}, D. 2006,
  \apjl, 646, L33

\bibitem[{{Yang} {et~al.}(2004){Yang}, {Zabludoff}, {Zaritsky}, {Lauer}, \&
  {Mihos}}]{Yang04}
{Yang}, Y., {Zabludoff}, A.~I., {Zaritsky}, D., {Lauer}, T.~R., \& {Mihos},
  J.~C. 2004, \apj, 607, 258

\bibitem[{{Yang} {et~al.}(2008){Yang}, {Zabludoff}, {Zaritsky}, \&
  {Mihos}}]{Yang08}
{Yang}, Y., {Zabludoff}, A.~I., {Zaritsky}, D., \& {Mihos}, J.~C. 2008, \apj,
  688, 945
  
\bibitem[{{Zabludoff} \& {Franx}(1993){Zabludoff} \& {Yang}}]{zab}
{Zabludoff}, A.~I. and {Franx}, M. 1993, \apj, 106, 1314

\bibitem[{{Zabludoff} \& {Mulchaey}(1998)}]{Zabludoff98}
{Zabludoff}, A.~I., \& {Mulchaey}, J.~S. 1998, \apjl, 498, L5+

\end{thebibliography}
\end{document}